\documentclass[twocolumn,aps,superscriptaddress,multicol,amsmath,amssymb]{revtex4-1}
\usepackage{amssymb}
\usepackage{amsmath}
\usepackage{graphicx}
\usepackage{epstopdf}
\usepackage{bm}
\usepackage{gensymb}
\usepackage{soul}
\usepackage{sidecap}
\usepackage[normalem]{ulem}
\usepackage{times}

\usepackage{graphicx}
\usepackage{epstopdf}
\usepackage{bm}
\usepackage{subfigure}
\usepackage{sidecap}

\usepackage{graphicx}
\usepackage{color}
\usepackage{epstopdf}
\usepackage{amssymb}
\usepackage{amsmath}

\usepackage{bm}
\graphicspath{{Figures/}}
\usepackage{subfigure}

\usepackage{lmodern}

\newcommand{\si}{\sigma}
\newcommand{\al}{\alpha}

\newcommand{\ga}{\gamma}
\newcommand{\tht}{\theta}

\newcommand{\lam}{\lambda}

\newcommand{\de}{\delta}
\newcommand{\De}{\Delta}
\newcommand{\Ga}{\Gamma}

\newcommand{\be}{\begin{equation}}
\newcommand{\ee}{\end{equation}}

\newcommand{\bea}{\begin{eqnarray}}
\newcommand{\eea}{\end{eqnarray}}
\newcommand{\bd}{\begin{displaymath}}
\newcommand{\ed}{\end{displaymath}}
\newcommand{\ba}{\begin{array}}
\newcommand{\ea}{\end{array}}
\newcommand{\bi}{\begin{itemize}}
\newcommand{\ei}{\end{itemize}}
\newcommand{\bc}{\begin{center}}
\newcommand{\ec}{\end{center}}
\newcommand{\bfl}{\begin{flushleft}}
\newcommand{\efl}{\end{flushleft}}
\newcommand{\bfr}{\begin{flushright}}
\newcommand{\efr}{\end{flushright}}

\newcommand{\bl}{\begin{aligned}}
\newcommand{\el}{\end{aligned}}

\newcommand{\hDe}{\hat{\Delta}}
\newcommand{\hJ}{\hat{J}}

\newcommand{\hchi}{\hat{\chi}}

\newcommand{\bJ}{\bar{J}}

\newcommand{\tf}{\tilde{f}}

\newcommand{\fs}{\frac{1}{2}}
\newcommand{\fss}{\frac{1}{\sqrt{2}}}

\newcommand{\om}{i\omega_n}

\newcommand{\ra}{\rangle}
\newcommand{\la}{\langle}

  \def\bq{{\bf q}}

 \def\bd{{\bf d}} \def\bS{{\bf S}} \def\bJ{{\bf J}}
 \def\bS{{\bf S}}

 \def\bJ{{\bf J}}

\def\={\!\!\!&=&\!\!\!}
\def\+{\!\!\!&&\!\!\!+~}
\def\-{\!\!\!&&\!\!\!-~}

\usepackage{color}
\usepackage[dvipsnames]{xcolor}
\usepackage{xcolor}
\def\BLU{\color{black}}  %

\usepackage[colorlinks=true,citecolor=blue]{hyperref}

\usepackage{hyperref,cleveref}

\begin{document}

\title{
Induced quantum magnetism in CEF singlet ground state models: Thermodynamics and  excitations}

\author{Peter Thalmeier}
\affiliation{Max Planck Institute for the Chemical Physics of Solids, 01187 Dresden, Germany}

\author{Alireza Akbari}
\affiliation{Asia Pacific Center for Theoretical Physics (APCTP), Pohang, Gyeongbuk, 790-784, Korea}
    %
%
\affiliation{Institut f\"ur Theoretische Physik III, Ruhr-Universit\"at Bochum, D-44801 Bochum, Germany}
\date{\today}

\begin{abstract}
We present a comparative investigation of singlet ground state induced magnetism for 
singlet, doublet and triplet excited CEF states of non-Kramers f-electrons
relevant primarily for Pr- and U- based compounds. This type
of magnetic order is of intrinsic quantum nature because it requires the 
superposition of singlet ground state with excited states due to non-diagonal matrix elements
of the effective intersite exchange to generate local moments. In contrast to conventional magnets the 
local moments and their ordering appear simultaneously at the transition temperature. It is finite only
if the control parameter proportional to the ratio of exchange strength to level splitting exceeds a critical
value marking the quantum critical point of the models. We determine the dependence of transition temperature, 
saturation moment, renormalised level splitting, specific heat jumps and low-temperature susceptibility as function of control parameter. Furthermore the temperature dependence of these quantities is calculated for control parameters above and below the quantum critical point and the distinction to conventional magnetism is discussed. In addition we investigate
the dynamical properties of the three models, deriving the magnetic exciton dispersion and their critical behaviour.
In particular the conditions for true and arrested soft-mode behaviour at the ordering wave vector are identified.
\end{abstract}


\maketitle

\section{Introduction}
\label{sec:introduction}

The majority of magnets with local moments can be understood within the context of semiclassical approach sufficiently 
below the ordering temperature $T_m$~\cite{white:83,majlis:07}. The formation of a collective moment and molecular field is due to the exchange coupling of stable local moments that exist already above $T_m$ and order spontaneously at this temperature, breaking time reversal and possibly spatial symmetries.
The ordered moment is treated as a classical variable and the quantum effects are included by considering the effect of  a  small number of bosonic excitations, i.e. magnons that reduce the ordered moment below the classical value by a certain amount. This reduction is moderate, e.g. in the three dimensional case, unless lower dimension and/or the effect of frustration leads to possibly divergent corrections heralding the breakdown of magnetic order and the appearance of a nonmagnetic 'spin liquid' ground state. Except for 1D magnetic chains such state can exist only in tiny parts of the exchange parameter space~\cite{schmidt:17a} and mostly the semiclassical picture is valid, even when the ordered moment reduction by quantum fluctuations and transition temperature suppression may be quite large~\cite{schmidt:17b}. \\

There is, however a class of magnetic materials with non-Kramers 4f- or 5f ions (with integer total angular momentum J) where the quasi-classical picture of ordering is per se invalid. This is the case when the crystalline electric field (CEF) splits the $(2J+1)$- fold degenerate J-multiplet  into a series of CEF multiplets such that the ground state $|0\rangle$ is a singlet which has no magnetic moment, i.e.  $\la 0|\bJ|0\ra=0$. In this situation the appearance of magnetic order is a true quantum effect, it can only happen if  at least one of the  \bJ~ components has non-diagonal matrix elements with nearby excited multiplets. Then a sufficiently strong intersite exchange may lead to the creation of an ordered moment by spontaneous formation of a new ground state which is a {\it superposition} of {\BLU singlet ground state and first excited multiplet state at an energy $\De$. If we assume an ideal situation where the latter is separated
from other much higher lying CEF levels and the transition temperature $T_m$  is small then in an intermediate temperature range $T_m<T<\De$ there will be no paramagnetic moment (just vanVleck terms) in the susceptibility. The moment then appears at $T_m$ as a collective ordered moment selfconsistently determined by the singlet-singlet mixing in the ordered phase which cannot be viewed quasiclassically as alignment of preexisting paramagnetic moments.}
The condition for the necessary strength of the intersite coupling is determined by a dimensionless control parameter composed of coupling strength, level splitting and the nondiagonal matrix elements (Eq.~(\ref{eq:control})). The critical value of the control parameter, equal to one, defines the quantum critical point (QCP) separating the paramagnetic (less than one) from the induced magnetic moment (larger than one) regime.\\

The thermodynamic signatures, as expressed by behaviour of critical temperature, saturation moment, specific heat anomalies and susceptibility are quite distinct from the quasi-classical magnets. Furthermore the dynamical characteristics show pronounced differences. In the latter coherent propagating magnons; i.e. quasi-classical precessing moments naturally can only appear in the {\it ordered} phase; requesting the presence of a molecular field. On the other hand in the induced  moment magnets collective 'magnetic exciton' modes are present already in the paramagnetic phase due to the possibility of dispersive inelastic CEF excitations between singlet ground state and excited multiplet. The dispersion of magnetic excitons is strongly temperature dependent controlled by the thermal population difference of CEF levels. Under suitable conditions it may turn into a soft mode at the incipient ordering wave vector as a precursor to spontaneous induced order. However, it is frequently arrested at a finite energy at the ordering temperature. Below the ordering temperature it reemerges as a strongly renormalised magnonic mode (possibly split into several branches) where the induced order parameter together with the molecular-field renormalised CEF level splitting determine the dispersion.
\\

This type of induced moment quantum magnetism, different from the common semiclassical variety has been known for some time but its possibility and distinction to quasi-classical magnetism is not commonly appreciated. It appears primarly 
in compounds with J=4 4f or 5f ions such as PrSb~\cite{mcwhan:79}, Pr$_3$Tl\cite{buyers:75,birgeneau:71}, PrCu$_2$\cite{kawarazaki:95}, PrNi\cite{savchenkov:19},  Pr- metal under pressure~\cite{jensen:91,birgeneau:72,cooper:72} but also Tb (J=6) compound TbSb~\cite{holden:74}. Furthermore 5f candidates for induced order are UGa$_2$~\cite{marino:23a}, UPd$_2$Al$_3$~\cite{grauel:92,mason:97,thalmeier:02} and URu$_2$Si$_2$\cite{sundermann:16} (and references cited therein), also Fe-substituted~\cite{marino:23b}. The latter example (when considered within the localised 5f scenario) and also the (Kramers ion) compound YbRu$_2$Ge$_2$ show that the induced order mechanism not only works for magnetism but also for multipolar~\cite{santini:94,haule:10} and quadrupolar~\cite{jeevan:06,takimoto:08} order, respectively. Finally another aspect of the singlet ground state induced magnetism has been discovered: For CEF singlet ground state f-electrons on suitable 2D lattices like honeycomb or kagome the magnetic excitons may develop a nontrivial topological character with non-vanishing Chern number which would entail the existence of magnetic excitonic edge modes in the paramagnetic state~\cite{akbari:23}. This possibility has previously only been considered for ferro- or antiferro- ordered 2D lattices~\cite{owerre:16,mcclarty:22}.\\

{\BLU
Another highly interesting aspect of induced moment magnetism is the possible influence of nuclear hyperfine coupling on the magnetic transition and order. The effects of hyperfine coupling and level splitting in thermodynamic properties of 4f systems appear in the $10^2$ mK regime \cite{steinke:13} in particular in Pr compounds since the $Pr^{141}$ isotope has the largest hyperfine coupling in the 4f series. When an induced moment system is accidentally close to 
the quantum critical point with zero or small ordering temperature the effect of the hyperfine coupling on the induced
order may become essential. A prominent example is Pr- metal which has slightly subcritical control parameter for purely 4f induced moment order, which may, however, be rapidly pushed above the critical parameter leading to finite transition temperature by applying uniaxial pressure \cite{jensen:87}. It has been suggested \cite{lindelof:75,jensen:79,moller:82,jensen:91} that as a result of hyperfine coupling between nuclear and 4f moments it nevertheless shows combined nuclear-electronic magnetic order around  50-60 mK already at ambient pressure. The importance of hyperfine coupling has also been proposed for  to explain the singlet ground state magnetism ($T_N\simeq 0.25$ K) of  Tb$_3$Ga$_5$O$_{12}$ and its excitations \cite{wawr:19}. Theoretical treatments of the combined effects of electronic exchange and nuclear hyperfine coupling were given e.g. in Refs.\cite{hamman:73,murao:79,jensen:91}. In this work we will not consider these further complications for the close-to critical induced moment magnets and focus only the purely electronic mechanism.}

Although the above mentioned compounds are known to have a singlet CEF ground state the type of the first excited state whether singlet, doublet or triplet (in cubic site symmetry only) is not always known with certainty, in particular in the U-compounds. However the quantitative and even qualitative aspects of induced moment magnetic order will be different in these three cases. Although they have been considered before in the references above there is no systematic comparison concerning their thermodynamic (specific heat, susceptibility etc.) or dynamic properties like distinct magnetic exciton dispersion and differences in temperature dependence and soft mode behaviour. In this work we 
undertake this effort to improve understanding of these physical properties of singlet ground state quantum magnets and  give a better foundation for possible judgement of experimentally observed properties. We will do this for three generally applicable models: The singlet-singlet, doublet or triplet models (SSM, SDM and STM respectively) which may be commonly realised in uniaxial (the two former) or cubic symmetry (the latter), see also Appendix \ref{app:CEF}.

Our approach is analytical based on molecular field (MFA) and random phase (RPA) approximations as far as it can be carried out. We emphasise as a unifying concept the central role of the control parameter of the three models defining the QCP and separating paramagnetic and magnetic phases. All physical properties will be expressed in terms of these control parameters. It has the further advantage that one can continuously approach the common magnetic order with
(quasi-) degenerate ground state by tuning the control parameter to large values far away from the QCP. In particular we shall focus on the temperature behaviour of susceptibility and specific heat in the ordered phase and on the concomitant control parameter dependence of ordered moment, ordering temperature and specific heat anomalies which show a clear distinction between the induced order quantum regime and the (asymptotic) quasi-classical regime.
Furthermore we calculate the magnetic excitation spectrum of the three models across the disordered and induced order regimes and demonstrate their striking differences. In particular we give an explicit demonstration how the 
simple soft mode picture is modified in the STM model leading to an arrested soft mode at the true transition temperature which is also of more general significance. This investigation is not focused on a specific material but rather on the comparative analysis of generally important singlet ground state CEF models. To keep the results from the three models distinct and avoid confusion we dedicate separate sections to them. This entails some redundancy but supports clarity of presented results.

\section{Models for induced quantum magnetism}
\label{sec:model}

\begin{figure}
\includegraphics[width=\linewidth]{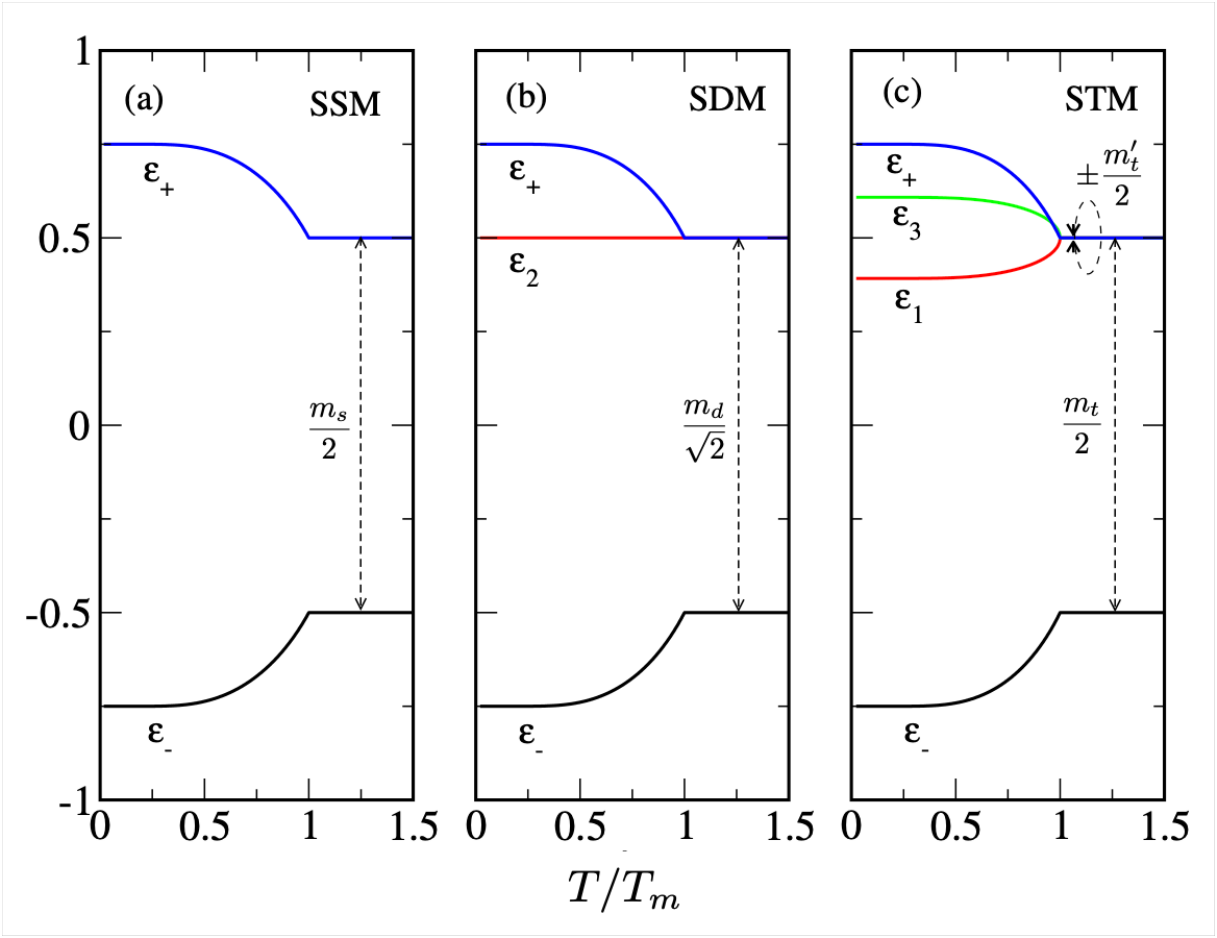} 
\caption{Behaviour of singlet ground and excited multiplet  level energies $\epsilon_\Ga-\frac{\De}{2}$ for the three models under the induced magnetic order (shifted by $\frac{\De}{2}$). Note that for SDM we use $\epsilon_1=\epsilon_-$ and $\epsilon_3=\epsilon_+$ in the related parts of the text. Here $\xi_{s,d,t} = 1.5$ corresponding to $T_m = 0.62, 0.51, 0.47$ consecutively. Nondiagonal matrix elements $m_{s,d,t}$ of moment operator necessary for induced order are indicated. In the STM a diagonal triplet matrix element $\pm m'_t=\pm 1$ occurs which leads to an enhancement of the transition temperature (Fig.~\ref{fig:Tm-xi}(a)) and an arrested mode softening (Fig.~\ref{fig:dispt-q}). Energy and temperature unit is $\De\equiv 1$ here and in all following graphs.}
\label{fig:levels-T}
\end{figure}

We consider three types of singlet- ground state level systems. The first two
are a singlet-singlet model (SSM) and a singlet-doublet model (SDM) frequently appropriate for uniaxial symmetry and a singlet-triplet model (STM) only possible for cubic symmetry. They correspond 
to simplified low energy CEF schemes consisting of just two levels whose splitting is considerably smaller than the excitation energies to the higher lying CEF states. Such systems are encountered
for even-numbered  f- electron shells with integer total angular momentum like Pr, U (J=4) , Tm, Tb (J=6) and Ho (J=8)  with examples given above. For the theoretical treatment of magnetic order and excitations it is a prerequisite to know the matrix structure of the angular momentum operators in the reduced SSM, SDM and STM level schemes. For uniaxial symmetries there are two possibilities, corresponding to Ising-type where only one operator, by convention $J_z$, or xy-type where two operators ($J_x,J_y$) have nonzero matrix elements between the  ground and excited CEF states of the reduced level scheme. Which one is realised depends on the Stevens parameters $B_n^m$  of the CEF potential and consequently the type of irreducible representations of low lying CEF states. Here we consider the Ising-type SSM, the xy-type SDM and the cubic STM cases. The CEF singlet ground state is denoted by $|0\rangle$. In the SSM the excited singlet is denoted by $|1\rangle$  and in the SDM the doublet states by $|1\sigma\rangle$ , $\sigma =\pm$, for STM see below. In all cases the CEF splitting energy is $\Delta$ (Fig.~\ref{fig:levels-T}).
The angular momentum operators for the xy-type  models within these low energy CEF subspaces have the general structure (in $|0\rangle, |1\rangle$ sequence):
$$
\mbox{Ising-type SSM:}
$$
\be
\bl
&
J_z
\!=\!
\frac{m_s}{2}
\left(
 \begin{array}{cc}
0& 1\\
1& 0
\end{array}
\right)=m_sS_x;
\label{eq:Jssm}
\el
\ee
where $\bJ$- operators refer to the free $|JM\ra$ states and $\bS$ are the pseudo-spin operators in the reduced subspace of  CEF states. Likewise in the singlet-doublet xy- type model the general form of \bJ~operators is given by  (in $|0\rangle, |1+\rangle, |1-\rangle $ sequence):
$$
\mbox{xy-type SDM:}
$$
\be
\bl
&
J_x=\frac{m_d}{\sqrt{2}}
\left(
 \begin{array}{ccc}
0& 1&1\\
1&0& 0\\
1&0&0
\end{array}
\right)=m_dS_x;
\\
&
J_y=\frac{m_d}{\sqrt{2}}
\left(
 \begin{array}{ccc}
0&i& -i\\
-i& 0&0\\
i&0&0
\end{array}
\right)=m_dS_y
,
\label{eq:Jsdm}
\el
\ee
 Here we defined $m_s$, $m_d$ in such a way that $S_x,S_y$ correspond to the canonical (pseudo-) spin matrices for $S=\fs,1$, respectively (with reordered sequence of states for the latter). The numerical values of $m_s, m_d$ are to be obtained from the diagonalisation of the full CEF Steven's Hamiltonian in a concrete case. \\
For the cubic STM we take as a model the most important $\Gamma_1-\Gamma_4$ level scheme of $J=4$ whose states are fully determined by symmetry and therefore do not depend on the CEF potential parameters. Since the cubic axes are equivalent we  restrict to the $J_z$ matrix where indices $n= 0,1-3$ correspond to the singlet $\Gamma_1$ ground state $|\psi_0\ra$ and triplet $\Gamma_4$ excited states $|\psi_{1-3}\ra$, respectively. It is given by
$$
\mbox{cubic- type STM:}
$$
\be
\bl
&
J_z=\frac{1}{2}
\left(
 \begin{array}{cccc}
0& 0&m_t &0\\
0&m'_t& 0& 0\\
m_t&0&0&0\\
0&0& 0& -m'_t\\
\end{array}
\right)=m_t(\fs\sigma_x)\oplus m'_t(\fs\tau_z)
,
\label{eq:Jstm}
\el
\ee
where $\sigma_x$ and $\tau_z$ are Pauli matrices in the $(0,2)$ and $(1,3)$ subspaces, respectively. For $J=4$ we have \cite{koga:06} $m_t=\frac{4}{3}\sqrt{15}=5.16$ for the nondiagonal element and $m'_t=1$ for the diagonal one. The ratio $m'_t:m_t=0.19$ controls to which extent the induced magnetic order is influenced by the excited magnetic $\Gamma_4$ triplet. This influence occurs only in the STM and has also important consequences for the softening behaviour of the magnetic exciton spectrum as a precursor to the induced order (Sec.~\ref{sec:exc-STM}).
This ratio is fixed by symmetry in the case of $J=4$ because $\Ga_1, \Ga_4$ representations occur only once. For higher $J=6,8$  they occur multiple times and therefore the ratio $m'_t:m_t$  depends on the CEF potential parameters which may then be considered as an additional variable parameter.\\

The effective intersite exchange interactions (mediated, e.g. by conduction electrons) together with the CEF potential
is described by the Hamiltonian
\be
H
\!=\!
\sum_{\Gamma,i}\epsilon_{\Gamma}|\Gamma,i\rangle\langle\Gamma,i|
-\!
\fs\sum_{\langle ij\rangle}J_{ij}\bJ_i\cdot\bJ_j
,
\label{eq:Ham}
\ee
where $\Gamma = 0,1$ or $0,(1+,1-)$ or $0,(1,2,3)$ labels the CEF states of the three models, respectively and $i,j$ denote the lattice sites. We restrict to only nearest neighbour  $\langle n.n. \rangle$ exchange interaction $J_{ij}=I_0$ for in-plane bonds and $J_{ij}=\kappa I_0$ for out-of plane bonds (along  c-axis), thus $\kappa$ controls the real-space anisotropy of the model and may be tuned continuously. We disregard possible spin-space anisotropies  of the exchange term to limit the number of parameters. The two CEF energy levels are given by  $\epsilon_0=0$ and $\epsilon_1$ , $\epsilon_{(1+,1-)}$ or $\epsilon_{(1-3)} =\Delta$, respectively. Frequently it will be convenient to use the more symmetrical shifted energy levels $\epsilon_\Gamma-\Delta/2$ (without introducing a new symbol). For simplicity we consider the localised  f electrons and CEF states on a simple tetragonal lattice. The corresponding (normalised) Fourier transformed n.n. exchange function is then given by (wave vectors in units of $\pi/a$ or $\pi/c$)
\be
\hat{I}(\bq)=
\frac{
{\rm sign}[I_0]
}
{
(2+\kappa)
}
(\cos q_x +\cos q_y +\kappa\cos q_z),
\label{eq:exfunc}
\ee
where $\hat{I}(\bq)=I(\bq)/I_e$ and the effective exchange coupling strength is given by $I_e=\frac{z}{3}|I_0|(2+\kappa)$ with $z=6$ denoting the n.n. coordination.  According to the sign of the exchange part in Eq.~(\ref{eq:Ham}) we use the convention that $I_0>0$ corresponds to FM coupling and $I_0<0$ to AF coupling.\\

Instead of the individual model parameters contained in the Hamiltonian of Eq.~(\ref{eq:Ham}) it will
turn out that in each of the three models one has only a single dimensionless control parameter given by 
\bea
\xi_{s,t}=\frac{m_{s,t}^2I_e}{2\Delta};
\quad
\xi_{d}=\frac{2m_{d}^2I_e}{\Delta},
\label{eq:control}
\eea
characterising the relative strength of effective intersite exchange $I_e$ and local CEF splitting $\Delta$. The slightly different definition of $\xi_d$ is due to two facts: In the SDM
{\it two} excited states are connected to the ground states and the prefactor of the matrix is chosen differently to comply with
pseudospin conventions for $S=1$. As it will turn out the above definition of the control parameters leads to identical quantum critical point $\xi^c_{s,d,t}=1$ for all three models {\BLU within the molecular field - RPA approach. This may
possibly change when considering the influence of interacting exciton modes beyond RPA  on the QCP position.} We will commonly suppress the $s,d,t$ indices when we talk about generic properties of all three models.
We will strive to express all relevant calculated quantities in terms of $\xi$ that controls the quantum phase transition from paramagnetic to induced moment phase. The conventional quasi-classical  magnetism in 3D is approached when $\xi\gg1$ and the CEF splitting becomes very small as compared to the exchange energy scale so that the first term in Eq.~(\ref{eq:Ham}) corresponds effectively to a quasi-degenerate multiplet.\\

The molecular field (MF) treatment of Eq.~(\ref{eq:Ham}) delivers the basic quantities of induced order parameter, transition temperature and level splitting and state superposition in the induced moment phase. Assuming that we either have possible ferromagnetic (FM) or antiferromagnetic (AFM) induced order (the only two possibilities for n.n. exchange) the effective molecular fields are given by $h_e^\lam=I_e\la J_\al\ra_{\bar{\lam}}$ ($\al= z$ for Ising- SSM and STM) and  we choose the moment direction for the xy- SDM along $\al=x$. Furthermore $\lam=A,B; \bar{\lam}=B,A$ is the sublattice index such that $h_e^\lam=h_e$ for FM and $h_e^A=-h_e^B=h_e$ for AFM cases. Then, e.g. for the SDM the molecular field Hamiltonian for each sublattice is given by
\be
H^\lam_{\rm MF}=\sum_i
[\sum_{\Gamma}\epsilon_{\Gamma}|\Gamma,i\rangle\langle\Gamma,i|
- h_e^\lam J_x(i)].
\label{eq:MFHam}
\ee
 After determination of the MF eigenvectors, energy eigenvalues and their thermal occupations  the induced moments $\la J_\alpha\ra$ may be computed selfconsistently  in the following.

\section{Thermodynamic properties: order parameter, specific heat and susceptibility}
\label{sec:thermo}

 The basic mechanism of induced order in singlet ground state systems  is in contrast to the quasi-classical case with (Kramers) degenerate  ground state as outlined in the Introduction.  In the three singlet models there are no preformed ground state moments, therefore magnetic order is only possible due to a nondiagonal quantum mixture of singlet ground state with the excited multiplet states due to inter-site exchange. This is a distinctly quantum mechanical origin of magnetic order caused by the superposition of {\it nonmagnetic }ground and excited states forming spontaneously a new magnetic ground state below $T_m$. It is natural for this mechanism to work that the intersite exchange driving the moment formation from the split states must be sufficiently large to overcome the CEF splitting stabilising the nonmagnetic state as controlled by the dimensionless parameters defined in Eq,~(\ref{eq:control}).
The technical treatment of the SSM, SDM and STM models will be rather similar. But for the sake of clarity of results we treat them in the separate following (sub-) sections.

\subsection{The Ising-type singlet-singlet model}

In the SSM the MF Hamiltonian of Eq.~(\ref{eq:MFHam})
has the sublattice- independent eigenvalues 
\be
\epsilon_\pm=\pm\fs\De_T=\pm\frac{\Delta}{2}[1+\gamma^2\la J_z\ra^2]^\fs
\label{eq:sMFlevel}
\ee
leading to a renormalised (dimensionless) singlet-singlet splitting in the ordered state according to 
\be
\bl
\hDe_T
&
=\frac{1}{\De}(\epsilon_+-\epsilon_-)=\frac{\Delta_T}{\De}=[1+\ga^2\la J_z\ra^2]^\fs;
\\
\ga
&
=
m_sI_e/\De=(2/m_s)\xi_s,
\el
\ee
which depends on temperature $T$ through the order parameter $\la J_z\ra$ as shown in Fig.~\ref{fig:levels-T}(a). 
The  latter also leads to a coherent mixing of ground and excited states {\BLU to the new eigenstates. Since the Hamiltonian is real symmetric they are given by the real orthogonal transformation}
\be
\bl
|\epsilon_+\ra
&
=
\cos\theta_\lam|1\ra - \sin\theta_\lam |0\ra,
\\
|\epsilon_-\ra
&
=\sin\theta_\lam|1\ra + \cos\theta_\lam |0\ra,
\label{eq:smagstate}
\el
\ee
where $\tan 2\theta_\lam= \gamma\la J_z\ra_\lam$ with $\theta_\lam=\theta$; $\la J_z\ra_\lam=\la J_z\ra $ for FM and $\theta_{A,B}=\pm\theta$;  $\la J_z\ra_{A,B}=\pm\la J_z\ra $ for AFM. The mixing of singlets is the essential mechanism to create a selfconsistent ground state  induced moment out of the singlet states  due to the nondiagonal dipolar matrix element $m_s$. This means that the mixing angle $\theta_\lam$ and $\la J_z\ra_\lam $ can only be simultaneously nonzero.

\subsubsection{Order parameter and transition temperature for SSM}
\label{sec:sOP}

\begin{figure}
\includegraphics[width=0.95\linewidth]{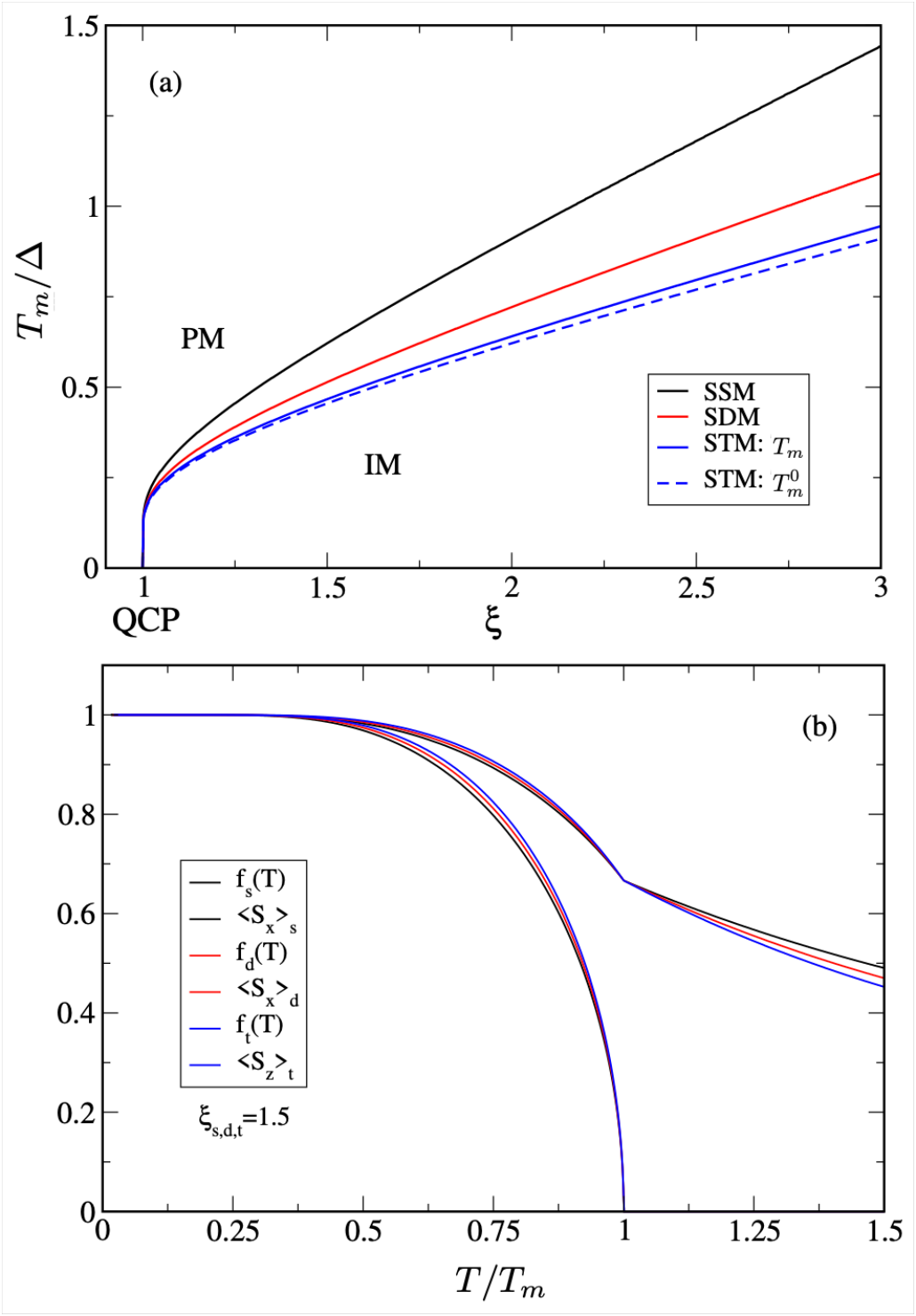}
\caption{(a) Critical temperature for induced magnetic order as function of control parameter $\xi$. The QCP value $\xi_c=1$ separates paramagnetic (PM, $\xi<1$) from induced moment (IM, $\xi>1$) regime. For STM both
exact $T_m$ (Eq.~(\ref{eq:tTm})) and approximate $T_m^0$ (Eq.~(\ref{eq:tTm0})) are shown. 
(b) Temperature dependence of population functions $f_{s,d,t}$ and corresponding normalised 
order parameter $\la S_x\ra_{s,d}$ and $\la S_z\ra_t$ for SSM , SDM and STM cases scaled by
$T_m = 0.62, 0.51, 0.47$, respectively.}
\label{fig:Tm-xi}
\end{figure}

From Eq.~(\ref{eq:smagstate}) the selfconsistent equation for the  order parameter induced by nondiagonal matrix element $m_s$ may be derived. Its temperature dependence originates from the thermal populations $p_\pm=Z^{-1}\exp(\mp\De_T/2T)$ of the two singlet states with energies $\epsilon_\pm$ where $Z=2\cosh(\De_T/2T)$ is the partition function. For the evaluation of the order parameter  $\la J_x\ra$ and later susceptibility $\hchi_{\al\al}$ components we need the form of $J_\alpha$ in Eq.~(\ref{eq:Jssm}) in the eigenvector basis of Eq.~(\ref{eq:smagstate}). 
as given by
\be
J_z=\frac{m_s}{2}
\left(
 \begin{array}{cc}
\sin 2\theta& \cos 2\theta\\
\cos2\theta& -\sin 2\theta
\end{array}
\right).
\label{eq:Jssmtrans}
\ee
The matrix elements for $J_z$  are determined by  $\cos 2\theta= [1+\gamma^2\la J_z\ra^2]^{-\fs}$ and  $\sin 2\theta=\gamma\la J_x\ra [1+\gamma^2\la J_z\ra^2]^{-\fs}$. For $\la J_z\ra=m_s\la S_x\ra$  we then can write
\be
\bl
\la J_z\ra_T
&=
\frac{m_s}{2}\frac{1}{\xi_s}[\xi_s^2f_s^2(T)-1]^\fs= \frac{m_s}{2}\frac{1}{\xi_s}[\hDe_T^2-1]^\fs,
\\
f_s(T)
&=\tanh[\bigl(\frac{\De_0}{2T}\bigr)f_s(T)]
=p_- -p_+
.
\label{eq:sOP}
\el
\ee
Here $\Delta_0=\xi_s\Delta=m_s^2I_e/2$ is the zero temperature splitting and $\hDe_T\equiv\De_T/\De=\xi_sf_s(T)$ the normalised T-dependent splitting. It is determined by the difference of thermal level occupations $f_s(T)$ for $T\leq T_m$ as obtained from the solution of the second equation for a given $\xi_s$. In the paramagnetic regime $(T>T_m)$ we simply have $f_s(T)\equiv f_s^0(T)=\tanh\frac{\De}{2T}$. This means $f_s(0)=1$ and the transition temperature $T_m$ is reached for $f_s(T_m)=\frac{1}{\xi_s}$, leading to
\be
\bl
T_m=\frac{\De}{2\tanh^{-1}(\frac{1}{\xi_s})}=
\left\{
\begin{array}{r l}
\frac{\De}{|\ln\frac{\delta}{2}|}\;\;\;& \xi_s=1+\delta\;\;\; (\delta\ll1)\\[0.3cm]
\fs\xi_s\De\;\;\;& \xi_s\gg 1
\label{eq:sTm}
\end{array}
\right.
,
\el
\ee
where the asymptotic limits are given to the right, the upper one corresponding to closeness to the QCP, the lower one approaching the conventional degenerate ground state magnetism. This means for a finite $T_m$ for induced quantum magnetic order one must have a control parameter $\xi_s>1$, i.e. according to Eq.~(\ref{eq:control}) an intersite exchange $I_e$ {\BLU which must be sufficiently large compared to the singlet-singlet CEF splitting as determined by
the nondiagonal matrix element according to $I_e/\De > 2/m_s^2$}. Therefore $\xi^c_s=1$ marks the quantum critical point (QCP) between paramagnetism $(\xi_s<\xi^c_s)$ and induced magnetic order  $(\xi_s>\xi^c_s)$. The normalized moment and the saturation moment $\la J_z\ra_0$ are given by 
\be
\bl
\la\hJ_z\ra_T
&
=\frac{\la J_z\ra_T}{\la J_z\ra_0}=\frac{[\xi_s^2f_s^2(T)-1]^\fs}{[\xi_s^2-1]^\fs},
\\
\la J_z\ra_0
&
=m_s\la S_x\ra_0
=
\frac{m_s}{2}\frac{1}{\xi_s}[\xi_s^2-1]^\fs
\\
&
=
\left\{
\begin{array}{r l}
m_s\sqrt{\frac{\delta}{2}}\;\;\;&  \xi_s=1+\delta\;\;\; (\delta \ll 1)\\[0.3cm]
\frac{m_s}{2}\;\;\;&  \xi_s\gg 1
\end{array}
\right..
\label{eq:momasym}
\el
\ee
Transition temperature and induced moment  are shown in Fig.~\ref{fig:Tm-xi} and discussed further in Sec.~\ref{sec:discussion}. It is also instructive to consider the {\BLU  ratio of  (pseudospin) moment to (normalized) transition temperature given
by 
\be
\bl
\frac{\la S_x\ra_0}{(T_m/\De)}
&=
\frac{1}{\xi_s}[\xi_s^2-1]^\fs\tanh^{-1}\bigl(\frac{1}{\xi_s}\bigr)
\\
&=
\left\{
\begin{array}{r l}
\sqrt{\frac{\delta}{2}}|\ln\frac{\delta}{2}|\;\;\;&  \xi_s=1+\delta\;\;\; (\delta \ll 1)\\[0.3cm]
\xi_s^{-1}\;\;\;&  \xi_s\gg 1
\end{array}
\right.
.
\el
\ee
}
Close to the QCP (upper limit) this ratio decreases steeply to zero \cite{marino:23b}, as opposed to the conventional magnetism limit $\xi_s\gg1$ where $\la S_x\ra_0\approx\fs$ corresponding to proportionality of transition temperature $T_m\sim I_e$ to the exchange strength
in this limit.

\subsubsection{Internal energy and specific heat for SSM}
\label{sec:sspec}

In the paramagnetic phase the internal energy of the SSM is simply $U(T)=-\frac{\De}{2}\tanh\bigl(\frac{\De}{2T}\bigr)$
leading to a Schottky specific heat (second line in Eq.~(\ref{eq:sCV})). In the ordered phase we have to include the direct contribution of the order parameter. 
Then the temperature dependent internal energy (per site) in MF approximation is given by
\be
U(T)=\sum_{\si}p_\si\epsilon_\si(T)+\frac{1}{2}I_e\la J_x\ra_T^2,
\label{eq:UT}
\ee
where the first part contains the MF energy levels of Eq.~(\ref{eq:sMFlevel}) and their thermal occupations. This may be evaluated by using the expressions derived before as 
\be
U(T)=
\left\{
\begin{array}{l l}
-\frac{\De}{4}(\xi_sf_s^2(T)+\frac{1}{\xi_s})\;\;\;& T\leq T_m\\[0.3cm]
-\frac{\De}{2}f_s^0(T)\;\;\;&  T>T_m
\end{array}
\right.
.
\label{eq:sUT}
\ee
Then $U(0)=-\frac{\De}{4}(\xi_s+\frac{1}{\xi_s})$ at zero temperature. For $\xi_s\rightarrow 1^+$ this approaches the paramagnetic ground state energy $\epsilon_0 =-\frac{\De}{2}$. The specific heat $C_V(T)=(\partial U(T)/\partial T)_V$ may now be calculated using Eq.~(\ref{eq:sOP}) leading to
\be
\bl
\left\{
\begin{array}{l l}
C_V(T)=\bigl(\frac{\De_T}{2T}\bigr)^2\bigl(\cosh^2\frac{\De_T}{2T}-\frac{\De_0}{2T}\bigr)^{-1}\;\;\;& T\leq T_m\\[0.3cm]
C_V^0(T)=\bigl(\frac{\De}{2T}\bigr)^2\cosh^{-2}\frac{\De}{2T}\;\;\;&  T>T_m
\end{array}
\right.
\label{eq:sCV}
\el
\ee
for magnetic and paramagnetic phases, respectively.
where $C_V^0(T)$ is the background of the Schottky anomaly for a two-level system (N=1 in Appendix \ref{app:schottky}). 
Since the specific heat for $T\leq T_m$ contains the effects of the order parameter slope $\partial\hDe(T)/\partial T$ which is discontinuous at $T_m$ (zero above and finite below) there is a jump $\de C_V=C_V^--C_V^+$ in the specific heat starting from the paramagnetic value $C_V^+=C^0_V(T_m^+)$ just above $T_m$ to the value  $C_V^-=C_V(T_m^-)$ just below. From the above equations we get
\be
\bl
C_V^+
&=
\frac{1}{\xi_s^2}(\xi_s^2-1)\bigl(\tanh^{-1}\frac{1}{\xi_s}\bigr)^2,
\\[0.2cm]
C_V^-
&=
\frac{\frac{1}{\xi_s^2}(\xi_s^2-1)\bigl(\tanh^{-1}\frac{1}{\xi_s}\bigr)^2}
{1-\frac{1}{\xi_s}(\xi_s^2-1)\tanh^{-1}\frac{1}{\xi_s}}=
\frac{C_V^+}{1-\lam_s C_V^+},
\\
\lam_s
&=\frac{\xi_s}{\tanh^{-1}\frac{1}{\xi_s}},
\\[0.2cm]
\de C_V
&=
C_V^+\frac{\lam_sC_V^+}{1-\lam_s C_V^+}
\\
&
\simeq
\left\{
\begin{array}{l l}
\fs\delta^2|\ln\frac{\delta}{2}|^3\;\;\;&  \xi_s=1+\delta\;\;\; (\delta \ll 1)\\[0.3cm]
\frac{3}{2};\;\;&  \xi_s\gg 1
\end{array}
\right.
,
\label{eq:sCV1}
\el
\ee
where we used $\cosh^2\bigl(\frac{\De}{2T_m}\bigr)=\xi_s^2/(\xi_s^2-1)$ and $\bigl(\frac{\De}{2T_m}\bigr)=\tanh^{-1}\frac{1}{\xi_s}$ from Eq.~(\ref{eq:sTm}). The relative jump size compared to the paramagnetic value is then 
\be
\bl
\frac{\delta C_V}{C_V^+}=\frac{\lam_sC_V^+}
{1-\lam_sC_V^+}\simeq
\left\{
\begin{array}{l l}
\delta|\ln\frac{\delta}{2}|\;\;\;&  \xi_s=1+\delta\;\;\; (\delta \ll 1)\\[0.3cm]
\frac{3}{2}\xi_s^2\;\;\;&  \xi_s\gg 1
\end{array}
\right.
.
\label{eq:sCV2}
\el
\ee
On approaching the QCP from above $(\xi_s\rightarrow 1^+)$ both paramagnetic (Schottky) value $C_V^+$ and jump value $\delta C_V$ vanish and their ratio also vanishes $\delta C_V/C_V^+\rightarrow 0$. For large $\xi_s$, approaching conventional magnetism the ratio increases $\sim \frac{3}{2}\xi_s^2$  because the Schottky value $C_V^+\simeq 1/\xi_s^2$ vanishes due to $\Delta/T_m\rightarrow 0$. This means the absolute jump value approaches $\de C_V\rightarrow \frac{3}{2}$ in this limit  which agrees with the value known from conventional magnets with twofold degenerate ground state~\cite{majlis:07}. The $\xi$-dependence of the specific heat and its jump for the three models is illustrated in Figs.~\ref{fig:CVT},\ref{fig:dCV-xi} and discussed in Sec.~\ref{sec:discussion}.

\subsubsection{The static homogeneous SSM susceptibility}
\label{sec:ssuscept}

It is particularly instructive to consider the temperature dependence of the susceptibility, {\BLU i.e. the response function of magnetic moments ${\bf m}=g_J\mu_B\bJ$} across the induced order phase transition both in the paramagnetic and induced order (FM or AFM) phases with field applied parallel to the moment direction $\la J_z\ra$. The general expression for the {\BLU normalised} single-ion susceptibility  $\hchi_{\al\al}^0=\chi_{\al\al}^0/(g_J\mu_B)^2$ $(\al=x,y,z)$ for split singlets only is given by~\cite{jensen:91} 
\bea
\bl
\hchi^0_{\al\al}
=
&
\sum_{n\neq m}|\la m| J_\al|n\ra |^2\frac{p_n-p_m}{\epsilon_m-\epsilon_n}
\\
&
+
\frac{1}{T}[\sum_n|\la n| J_\al|n\ra |^2p_n-\la J_\al\ra^2]
\\
=& \;
\hchi^{0VV}_{\al\al}+\hchi^{0C}_{\al\al},
\label{eq:baresus}
\el
\eea
where the first part is the vanVleck term (VV) and the second a pseudo-Curie (C) term active at  temperatures comparable to the CEF splitting due to diagonal matrix elements. {\BLU These may already be present in the paramagnetic state as in STM or they may be induced in the ordered phase by the rotation to new eigenvectors as in SSM and SDM. In any case the second contribution vanishes, however, exponentially for low temperatures.} The collective (RPA) susceptibilities are then obtained via
\bea
\hchi_{\al\al}(T)=\frac{\hchi^0_{\al\al}(T)}{1\mp I_e\hchi^0_{\al\al}(T)}
\label{eq:sus}
\eea
with $I_e=\frac{z}{3}|I_0|(2+\kappa)$. Here the upper (lower) sign correspond to FM $(I_0>0)$ or AFM $(I_0<0)$ cases, respectively.
For the evaluation of susceptibilities in the ordered phase we need the forms of $J_z$ in Eq.~(\ref{eq:Jssm}) in the eigenvector basis according to Eqs.~(\ref{eq:smagstate},\ref{eq:Jssmtrans}). Then, using the Eqs.~(\ref{eq:baresus},\ref{eq:sus}) we obtain in the paramagnetic state:
$$T>T_m:$$
\be
\bl
&
\hchi^{VV0}_{zz}(T)
=
\frac{1}{I_e}\xi_sf_s^0(T);
\\
&
\hchi^{C0}_{zz}(T)=0;
\\
&
\hchi_{zz}(T)=
\frac{1}{I_e}\frac{\xi_sf_s^0(T)}{1\mp I_ef_s^0(T)}
;
\\
&
\hchi_{zz}(T_m^+)=
\left\{
\begin{array}{r l}
\rightarrow\infty& (-): {\rm FM }\\[0.3cm]
\frac{1}{2I_e}& (+):  {\rm AFM}
\end{array}
\right.;\;\;\;
\\
&
\hchi_{zz}(T\gg T_m)
\approx \frac{m_s^2}{4}\frac{1}{T}
.
\label{eq:ssuspara}
\el
\ee
In the induced moment phase we obtain for the $zz$ component, using  the abbreviation $f_s=f_s(T)$ and recalling that $\De_0=\xi_s\De$ and $(\De_0/2T_m)=\xi_s\tanh^{-1}\frac{1}{\xi_s}$:
$$T\leq T_m: $$
\be
\bl
& \hchi^{VV0}_{zz}(T)
=
\frac{1}{I_e}\frac{1}{\xi^2_sf^2_s};\;\;\; 
\\
 &\hchi^{C0}_{zz}(T)
 =
 \frac{1}{I_e}\bigl(\frac{\De_0}{2T}\bigr)\frac{(\xi^2_sf^2_s-1)(1-f^2_s)}{\xi^2_sf^2_s} ;
 \\[0.3cm]
&\hchi_{zz}(T)
=
\\
&\hspace{0.1cm}
\frac{1}{I_e}\frac{1+\bigl(\frac{\De_0}{2T}\bigr)(\xi^2_sf^2_s-1)(1-f^2_s)}
{\xi^2_sf^2_s\mp[1+\bigl(\frac{\De_0}{2T}\bigr)(\xi^2_sf^2_s-1)(1-f^2_s)]};
\\[0.2cm]
 &\hchi_{zz}(T_m^-)=
\left\{
\begin{array}{r l}
\rightarrow\infty& (-): {\rm FM }\\[0.3cm]
\frac{1}{2I_e}& (+): {\rm AFM}
\end{array}
\right.
;
\\&
\hchi_{zz}(0)=
\left\{
\begin{array}{r l}
\frac{1}{I_e}\frac{1}{\xi_s^2-1}& (-): {\rm FM}
 \\[0.3cm]
\frac{1}{I_e}\frac{1}{\xi_s^2+1}& (+): {\rm AFM}
\end{array}
\right.
.
\label{eq:ssusmag}
\el
\ee

The complicated algebraic structure in this case is mostly due to the second pseudo-Curie term. As $f_s(0)=1$ and $f_s(T_m)=1/\xi_s$ we notice that it vanishes both for $T=0$ and $T=T_m$ but is nonzero in between. The total susceptibility  $\hchi_{zz}(0)$  is also non-vanishing for T=0 because the induced saturation moment $\la J_z\ra_0$ (Eq.~(\ref{eq:momasym})) is not fully polarised for moderate $\xi_s$. This is a most characteristic difference to conventional magnets with degenerate ground state  where the fully developed saturation moment $\la J_z\ra_0=m_sS\; (S=\fs)$ can no longer be polarised and therefore $\hchi_{zz}(0)=0$. The induced moment case approaches that conventional limit for $\xi_s\gg 1$. These results are presented in Fig.~\ref{fig:susz-T} and discussed further in Sec.~\ref{sec:discussion}.

\subsection{The xy-type  singlet-doublet model}
\label{sec:xy-singlet}

The SDM model consisting of a singlet ground state $|0\ra$ with $\epsilon_0=0$ and a doublet $|1\pm\ra$ at $\epsilon_{1\pm}=\De$ is defined by the Hamiltonian of Eq.~(\ref{eq:Ham}). When convenient we also use shifted values $\epsilon_\Gamma-\frac{\De}{2}$ (without introducing new symbols). The MF treatment according to Eq.~(\ref{eq:MFHam}) leads to the split three-singlet eigenvalues given by
\be
\epsilon_{1,3}=\frac{\De}{2}
\Big[
1\mp[1+2\bigl(\frac{2\ga\la J_x\ra}{\De}\bigr)^2
\Big]^\fs;\;\;\;
\epsilon_2=\Delta
,
\label{eq:dCEFlevel}
\ee
now with $\gamma=\frac{m_d}{\sqrt{2}}I_e$.
Although the singlet  ground state mixes with both of the excited states only the $|\epsilon_{1,3}\ra$ show repulsion whereas the energy  of $|\epsilon_2\ra$ remains unaffected because the corresponding eigenstate is the antisymmetric combination $\frac{1}{\sqrt{2}}(|1+\ra-|1-\ra)$ as shown below. Altogether the orthonormal eigenstates in row order of increasing energies $\epsilon_{1,2,3}$ are given by the columns of the unitary $(U^\dagger U=1)$ matrix
\be
\bl
U=
\left(
 \begin{array}{ccc}
\sin\tht_1& 0&\sin\tht_3\\
-\fss\cos\tht_1&\fss& -\fss\cos\tht_3\\
-\fss\cos\tht_1&-\fss&-\fss\cos\tht_3
\end{array}
\right),
\el
\ee
where the matrix elements expressed by the mixing angles $\theta_i$ are given by (i=1,3)
\be
\bl
\cos\tht_i
=& 
\Big[1+\fs
\Bigl(\frac{2\ga\la S_x\ra}{\epsilon_i}
\Bigr)^2
\Big]^{-\fs},
\\
\sin\tht_i
=&\fss
\Big(\frac{2\ga\la S_x\ra}{\epsilon_i}\Big)
\Big[1+\fs
\Big(\frac{2\ga\la S_x\ra}{\epsilon_i}
\Big)^2
\Big]^{-\fs}.
\el
\ee
 From the structure of $U$ we can see that $|\epsilon_2\ra$ is just the antisymmetric combination of the original degenerate excited doublet components with unchanged $\epsilon_2=\Delta$ whereas $|\epsilon_{1,3}\ra$ have components of all states mixed together and their energies $\epsilon_{1,3}$ repel symmetrically leading to three singlets in the ordered state (Fig.~\ref{fig:levels-T}(b)).

\begin{figure}
\includegraphics[width=0.95\linewidth]{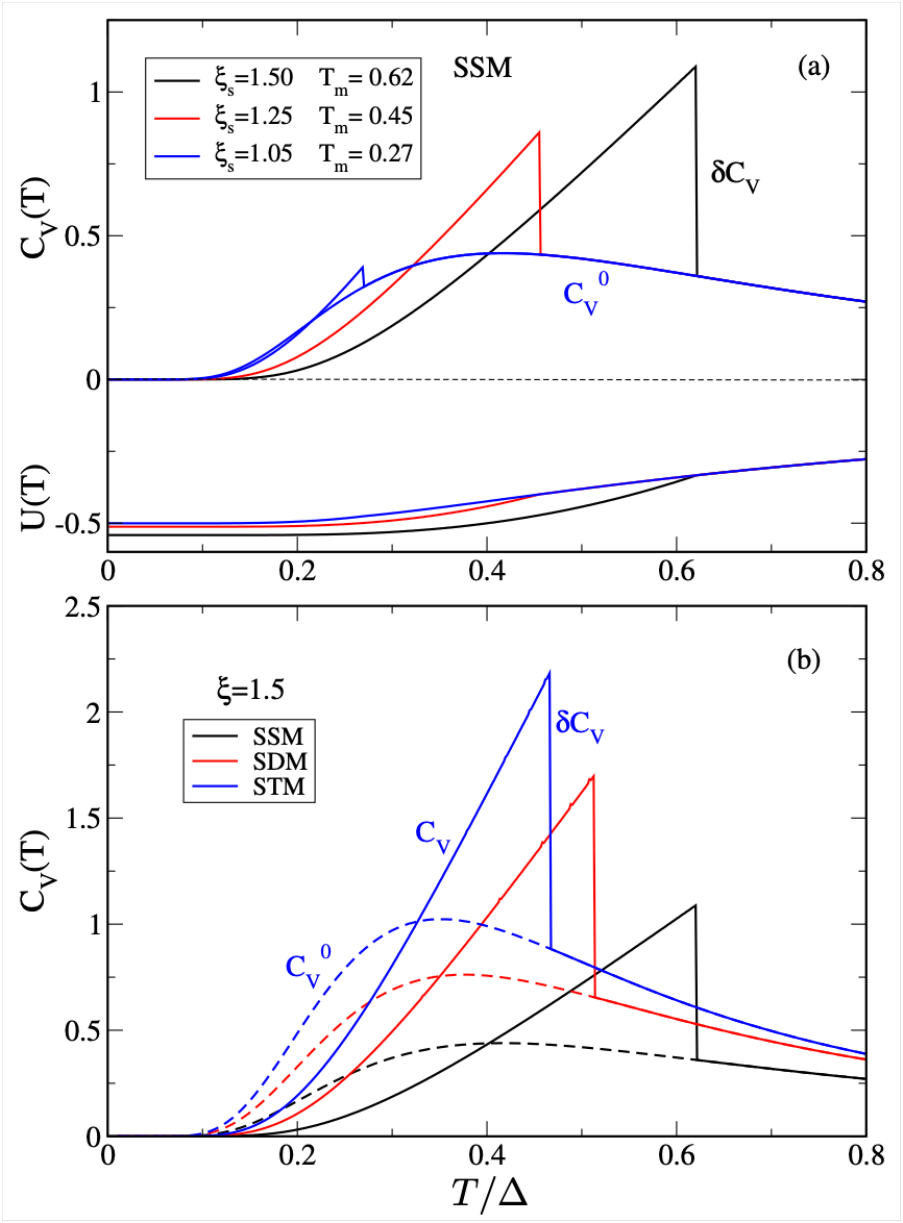}
\caption{Specific heat in units of $k_B$/site. Note in this figure $T$ is normalised to the CEF splitting $\De$. (a) Temperature dependences of internal energy $U$ and specific heat $C_V$  for the SSM. On approaching the QCP for $\xi\rightarrow 1^+$ the jump $\de C_V$ due to induced order shifts downwards with $T_m$ and becomes progressively smaller. It is superposed on the background Schottky peak due to the CEF splitting.
(b) Complementary specific heat for the three models with same control parameter $\xi=1.5$ in each case. While the jump increases the transition temperature decreases with $T_m = 0.62, 0.51, 0.47$ consecutively. The $\xi-$ dependence of the jump $\delta C_V$ is shown in Fig.~\ref{fig:dCV-xi}.}
\label{fig:CVT}
\end{figure}
\begin{figure}
\includegraphics[width=0.95\linewidth] {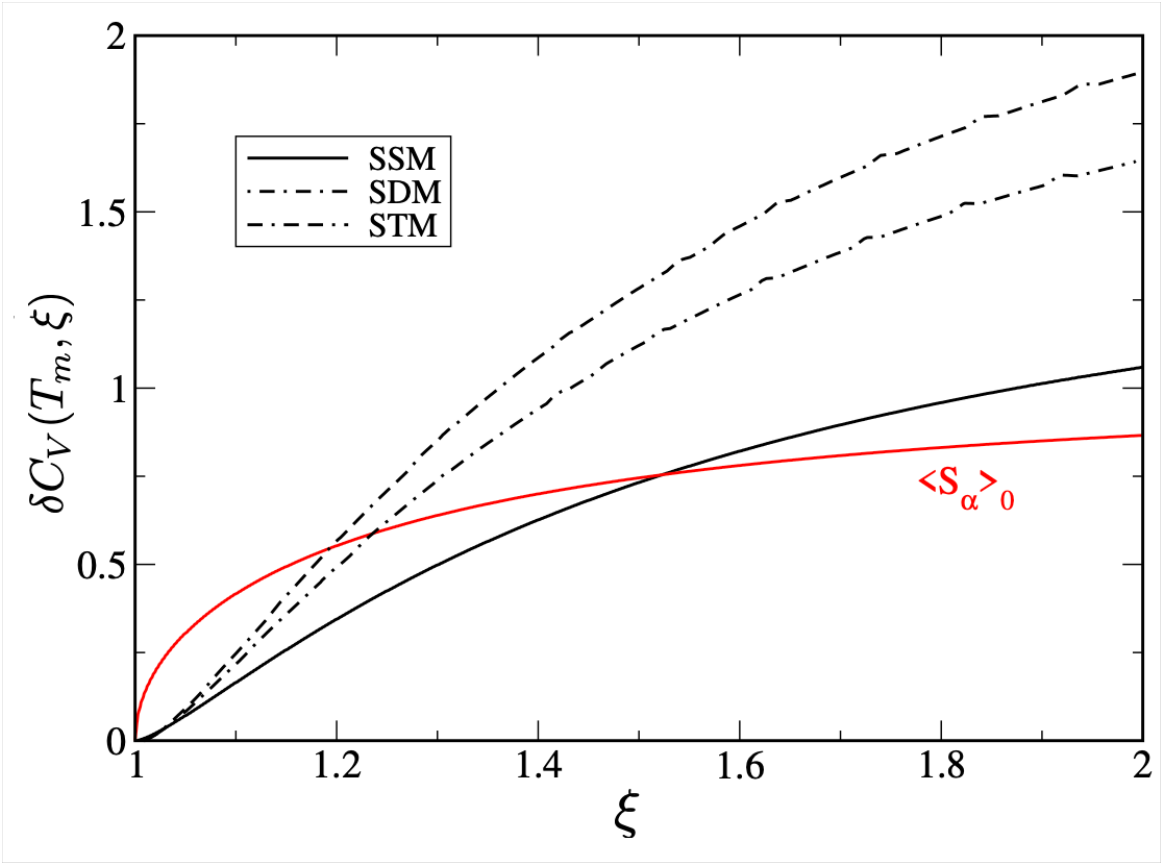}
\caption{Dependence of absolute specific heat jump $\de C_V=C_V^- -C_V^+$ at $T_m$ on the control parameter of the three models evolving in accordance with the size of the normalised  induced saturation order parameter $\la S_\al\ra_0$. Here  $\al=x$ for SSM and SDM and $\alpha=z$ for STM are identical. Close to the QCP with $\xi=1+\de$ $(\de\ll1)$ the moment vanishes $\sim(\de/2)^\fs$ with singular slope while the specific heat jump vanishes more gradually according to $\de C_V\sim\fs\de^2|\ln\frac{\de}{2}|^3$.}
\label{fig:dCV-xi}
\end{figure}

\subsubsection{Order parameter and transition temperature for SDM}
\label{sec:dOP}

For deriving the selfconsistency equation for induced moment $\la J_x\ra$ and later the susceptibilities $\hchi_{\al\al}$ we need again the $J_\al$ matrices in the eigenvector basis given by $U$, i.e. we have to transform
$J_\al\rightarrow U^\dagger J_\al U$. After some algebra this results in 
$$
\mbox{SDM:}
$$
\be
\bl
&
J_x=\frac{m_d}{\sqrt{2}}
\left(
 \begin{array}{ccc}
m_0& 0&m_1\\
0&0& 0\\
m_1&0&-m_0
\end{array}
\right)
;
\\&
J_y=\frac{m_d}{\sqrt{2}}
\left(
 \begin{array}{ccc}
0&im_2& 0\\
-im_2& 0&-im'_2\\
0&im'_2&0
\end{array}
\right),
\label{eq:Jsdmtrans}
\el
\ee
%
where the transformed matrix elements are combinations of the sines and cosines of mixing angles $\theta_{1,3}$ in U.
They may be transformed to the simple expressions
\be
\bl
J_x:\;\;\; m_0&=&\frac{\sqrt{2}}{\hDe_T}(1-\hDe_T^2)^\fs;\;\;\;
m_1=-\frac{\sqrt{2}}{\hDe_T},
\\
J_y:\;\;\;m_2&=&(1+\frac{1}{\hDe_T});\;\;\;
m'_2=(1-\frac{1}{\hDe_T}),
\label{eq:Jsdm2}
\el
\ee
%
%
%
where $\hDe_T=\xi_df_d(T)$ is defined below. They fulfil the sum rules $\fs(m_0^2+m_1^2)=\fs(m_2^2+m^{'2}_2)=1$. Note that $J_x$ matrix elements are nonzero  where those of $J_y$ vanish and vice versa. Using Eq.~(\ref{eq:Jsdmtrans}) leads to an order parameter
given by
\be
\la J_x\ra_T
=\frac{m_d}{\sqrt{2}}\frac{1}{\xi_d}[\xi_s^2f_d^2(T)-1]^\fs= \frac{m_d}{\sqrt{2}}\frac{1}{\xi_d}[\hDe_T^2-1]^\fs
.
\label{eq:dOP1}
\ee
Similar to Eq.~(\ref{eq:sOP}) here $\Delta_0=\xi_d\Delta$ is the renormalised zero temperature splitting and $\hDe_T=\xi_df_d(T)$. The selfconsistent equation for the temperature dependence of the occupation difference  $p_1-p_3=f_d(T)$ for $T\leq T_m$ is now more complicated due to the thermal population of the $|\epsilon_2\ra$ state at the unrenormalised energy $\Delta$ resulting from the lower state of the excited doublet split by $\la J_x\ra$ (Fig.~\ref{fig:levels-T}(b)). We obtain
\be
\bl
f_d(T)
=
&
 \frac{\tanh
 [\bigl(\frac{\De_0}{2T}\bigr)f_d(T)]
 }{1+\tf_d(T)}
 \\
 =
&
\frac{\sinh[\bigl(\frac{\De_0}{2T}\bigr)f_d(T)
\bigr]}
{\fs\exp(-\frac{\De}{2T})+\cosh[\bigl(\frac{\De_0}{2T}\bigr)f_d(T)
]
}
\label{eq:dOP2}
\el
\ee
with $\tf_d(T)=\fs\exp(-\frac{\De}{2T})/\cosh[\bigl(\frac{\De_0}{2T}\bigr)f_d(T)]$. If we would (wrongly) ignore the lower doublet state (setting $\tf_d=0$) the expression would become formally identical to that of the SSM case in Eq.~(\ref{eq:sOP}). Thus $\tf_d$ describes the correction due to the doublet nature of the excited CEF level.  In the paramagnetic regime we simply have to replace $\De_0f_d\rightarrow \De$ and get
$f_d^0(T)=2\tanh(\frac{\De}{2T})/[3-\tanh(\frac{\De}{2T})]$. As for the SSM the critical temperature for the induced moment $\la J_x\ra$ to appear is given by $f_d(T_m) 
= \frac{1}{\xi_d}$ with the solution
\be
T_m=\frac{\De}{2\tanh^{-1}\bigl(\frac{3}{1+2\xi_d}\bigr)}
\label{eq:dTm}
\ee
that depends in a different manner on the control parameter $\xi_d$ as compared to the singlet model (Eq.~(\ref{eq:sTm})). This is again caused by the finite thermal population of the lower doublet level at $T_m$. Nevertheless the QCP where $T_m$ vanishes, is identical to the SSM, given by $\xi^c_d=1$ (determined by the request that $3/(1+2\xi_d)<1$). Furthermore the normalised T-dependent and saturation moments for the SDM are similarly given by 
\be
\bl
\la\hJ_x\ra_T
&=
\frac{\la J_x\ra_T}{\la J_x\ra_0}
=
\frac{[\xi_d^2f_d^2(T)-1]^\fs}{[\xi_d^2-1]^\fs},
\\
\la J_x\ra_0
&=
m_d\la S_x\ra_0
=\frac{m_d}{\sqrt{2}}\frac{1}{\xi_d}[\xi_d^2-1]^\fs
\\
&=
\left\{
\begin{array}{r l}
\frac{m_d}{\sqrt{2}}\sqrt{\frac{\delta}{2}}\;\;\;&  \xi_d=1+\delta\;\;\; (\delta \ll 1)\\[0.3cm]
\frac{m_d}{\sqrt{2}}\;\;\;&  \xi_s\gg 1
\end{array}
\right.
.
\label{eq:dOP}
\el
\ee

\subsubsection{Internal energy and specific heat for SDM}
\label{sec:dspec}

The internal energy of the paramagnetic SDM model is $U(T)=-\frac{\De}{2}(3\tanh\bigl(\frac{\De}{2T}\bigr)-1)/(3-\tanh\bigl(\frac{\De}{2T}\bigr)$, which results in a correspondingly modified SDM Schottky-type specific heat (see N=2 case in Appendix \ref{app:schottky})
\be
C_V(T)=2\frac{4\bigl(\frac{\De}{2T}\bigr)^2}
{[3\cosh\bigl(\frac{\De}{2T}\bigr)-\sinh\bigl(\frac{\De}{2T}\bigr)]^2}
.
\label{eq:dCV}
\ee
For the magnetically ordered case it is determined by an expression corresponding to Eq.~(\ref{eq:UT}) for the SSM, now summing over three MF levels and occupations. Using the shifted eigenvalues $\epsilon_\Gamma-\frac{\De}{2}$ this leads again to
\be
U(T)=-\frac{\De}{4}\Bigl[\bigl(\xi_df_d^2(T)+\frac{1}{\xi_d}\bigr) -2\tf_d(T)\bigl(1+\tf_d(T)\bigr)^{-1} \Bigr].
\label{eq:dUT}
\ee
The first part is like SSM case and the second one corrects for the additional doublet level. For  zero temperature we again have $U(0)=-\frac{\De}{4}(\xi_d+\frac{1}{\xi_d})$. Due to the population effect of the additional level analytical derivation and discussion of $C_V(T)=(\partial U(T)/\partial T)_V$ is no longer reasonably feasible, considering the complicated MF equation Eq.~(\ref{eq:dOP2}) for $f_d(T)$. Therefore in the ordered phase, once this function has been determined the specific heat is obtained by numerical differentiation of $U(T)$.

\subsubsection{The static homogeneous SDM susceptibility}
\label{sec:dsuscept}

The basic expressions for single-ion and collective susceptibilities in Eqs.~(\ref{eq:baresus},\ref{eq:sus}), now with level energies and  matrix elements  corresponding to the SDM (Eqs.~(\ref{eq:dCEFlevel},\ref{eq:Jsdm})) are used to calculate these quantities. For the paramagnetic state the expressions for the susceptibility components are completely
equivalent to Eq.~(\ref{eq:ssuspara}) with the replacement $(m_s, \xi_s, f^0_s) \rightarrow (m_d, \xi_d, f^0_d)$. For the magnetic phase we obtain different expressions
$$
(xx): T\leq T_m:
$$
\be
\bl
&\hchi^{VV0}_{xx}(T)
=\frac{1}{I_e}\frac{1}{\xi^2_df^2_d};
\\
&
 \hchi^{C0}_{xx}(T)=\frac{1}{I_e}\bigl(\frac{\De_0}{2T}\bigr)\frac{(\xi^2_df^2_d-1)((1+\tf_d)^{-1}-f^2_d)}{\xi^2_df^2_d} ;
 \\[0.3cm]
&\hchi_{xx}(T)
=
\\
&
\hspace{0.5cm}
\frac{1}{I_e}\frac{1+\bigl(\frac{\De_0}{2T}\bigr)(\xi^2_df^2_d-1)((1+\tf_d)^{-1}-f^2_d)}
{\xi^2_df^2_d\mp[1+\bigl(\frac{\De_0}{2T}\bigr)(\xi^2_df^2_d-1)((1+\tf_d)^{-1}-f^2_d)]};
\\[0.2cm]
&\hchi_{xx}(T_m^-)
=
\left\{
\begin{array}{r l}
\rightarrow\infty& (-): {\rm FM} \\[0.3cm]
\frac{1}{2I_e}& (+): {\rm AFM}
\end{array}
\right.
;
\\
&
\hchi_{xx}(0)=
\left\{
\begin{array}{r l}
\frac{1}{I_e}\frac{1}{\xi_d^2-1}& (-): {\rm FM} \\[0.3cm]
\frac{1}{I_e}\frac{1}{\xi_d^2+1}& (+): {\rm AFM}
\end{array}
\right.
.
\label{eq:dsusmag}
\el
\ee
Obviously the values for $T=0,T_m$ are  equivalent to the SSM case  but the T-dependence in between is
modified by the presence of the $(1+\tf_d)^{-1}$ terms due to the lower doublet state. 

 For the transverse (yy) component the Curie terms are absent since $J_y$ is unchanged in the eigenvector basis and therefore has no diagonal matrix elements. And then we get the simple result
\be
\bl
\hchi_{yy}(T)=
\left\{
\begin{array}{r l}
\rightarrow \infty& (-):  {\rm  FM} \\[0.3cm]
\frac{1}{2I_e}& (+):  {\rm AFM}
\end{array}
\right.
.
\label{eq:yyssusmag}
\el
\ee
In the FM case the (yy) component diverges for all $T<T_m$ because there is no in-plane exchange anisotropy
and the ordered moment direction can be rotated from $x$ to any other direction by arbitrary small field. This is connected with the existence of a Goldstone mode for the whole ordered region (Sec.~\ref{sec:exc-SSM}). In the AFM phase the two sublattice moments tilt in addition to their rotation ($\perp$ to the field) leading to the finite susceptibility which is constant throughout the ordered phase as in conventional magnets. The transverse $(yy)$ susceptibility is then simply a constant $\hchi_{yy}=1/2I_e$  for the AF case. The comparison of susceptibilities for all models is presented in Fig.~\ref{fig:susz-T} for various positions of the control parameters with respect to the QCP. For a further discussion see Sec.~\ref{sec:discussion}.

\subsection{The cubic singlet-triplet model}
\label{sec:triplet}

In the STM model the three cubic axes are equivalent and for convenience we choose $\langle J_z\rangle$ direction for the induced magnetic moment. The corresponding molecular field Hamiltonian with  symmetric shifted CEF $\Gamma_1, \Gamma_4$ levels at $(-\frac{\De}{2},\frac{\De}{2})$ may be written as
\be
H_{\rm MF}^\lam=\sum_i[\frac{\De}{2}(P_4(i)-P_1(i))-h_e^\lam J_z(i)].
\ee
Here $P_{1,4}$ are projectors to $\Gamma_1$ singlet and $\Gamma_4$ triplet subspaces spanned by $|\psi_0\ra$ and $|\psi_{1,2,3}\ra$ states, respectively and $h_e^\lam=I_e\la J_z\ra$ is the molecular field. Using Eq.~(\ref{eq:Jstm}) the eigenvalues and -vectors of $H_{\rm MF}$ may be obtained easily because only $|\psi_{0,2}\ra$ are mixed by a nondiagonal matrix element (Eq.~(\ref{eq:Jstm})). We obtain the four level scheme totally split by the molecular field:
\be
\epsilon_\pm=\pm\frac{\De}{2}(1+\gamma^2\la J_z\ra^2)^\fs;\;\;\; \epsilon_{1,3}=\frac{1}{2}(\De\mp\delta_{T}),
\ee
where $\gamma=m_tI_e/\De=\frac{2}{m_t}\xi_t$. The symmetric normalised splittings of $\Gamma_1-\Gamma_4$ $|\psi_{0,2}\ra$ and $\Gamma_4$ $|\psi_{1,3}\ra$ states and their temperature dependence are then given by $\hDe_T=(1+\ga^2\la J_z\ra^2)^\fs$ and  $\hat{\delta}_T=\delta_T/\De=I_e\la J_z\ra/\De$, respectively (Fig.~\ref{fig:levels-T}(c)). Since only the  $|\psi_{0,2}\ra$ mix we have again

\be
\bl
\label{eq:tmagstate}
|\epsilon_+\ra&=\cos\theta_\lam|\psi_2\ra - \sin\theta_\lam |\psi_0\ra,
\\
|\epsilon_-\ra&=\sin\theta_\lam|\psi_2\ra + \cos\theta_\lam |\psi_0\ra,
\el
\ee
with $\tan 2\theta_\lam=\gamma\la J_z\ra_\lam$. The wave functions of split $\epsilon_{1,3}$ triplet levels are unchanged.
In the ordered phase the moment operator is then given by
\bea
\bl
J_z=\frac{1}{2}
\left(
 \begin{array}{cccc}
-m_t\sin 2\theta& 0&m_t\cos2\theta &0\\
0&1& 0& 0\\
m_t\cos 2\theta&0&m_t\sin 2\theta&0\\
0&0& 0& -1\\
\end{array}
\right).
\label{eq:Jstmfield}
\el
\eea
The order parameter is determined by the thermal trace of the diagonal elements and all states will contribute 
to $\la J_z\ra$, but only the elements $\sim\sin\theta$ are induced by the order itself, whereas the constant diagonal  elements are regular but thermally activated contributions.

\subsubsection{Order parameter and transition temperature for STM}

 Using the diagonal elements of $J_z$ for the eigenstates  and the split level energies the MF equation for 
the order parameter $\la J_z\ra$ is given by 
\bea
\la J_z\ra=\frac{\xi_t}{\hDe_T}\la J_z\ra P_a(T)+P_b(T)
\label{eq:tOP}
\eea
with the contributions resulting from the induced moment (a) and splitting of the excited triplet (b), respectively.
Here $P_a,P_b$ the differences of the thermal level populations $p_i(T)=Z^{-1}\exp(-\epsilon_i/T)$ $(i=\pm,1,3)$ with the partition function given by 
\be
\bl
Z=
&
2\bigl[\cosh(\frac{\epsilon}{2T})+\exp(-\frac{\De}{2T})\cosh(\frac{\delta_{s}}{2T})\bigr]
\\
&
\rightarrow
Z_0=2(2\cosh\frac{\De}{2T}-\sinh\frac{\De}{2T}).
\label{eq:tpart}
\el
\ee
Then we obtain, with the expressions to the right of the arrows corresponding to the paramagnetic state ($\hat{\delta}_T=0, \De_T=\De$):
\be
\bl
P_a(T)
=
\;
&
p_- -p_+=\frac{\tanh\frac{\De_T}{2T}}{1+g(T)}\rightarrow
\frac{\tanh\frac{\De}{2T}}{2-\tanh\frac{\De}{2T}}
,
\\
P_b(T)
=
\;
&
p_1-p_3=\frac{g(T)\tanh\frac{\de_T}{2T}}{1+g(T)}\rightarrow 0
,
\\
g(T)
=
\;
&\exp(-\frac{\De}{2T})\frac{\cosh\frac{\delta_T}{2T}}{\cosh\frac{\De_T}{2T}}\rightarrow 
1-\tanh(\frac{\De}{2T})
.
\label{eq:tpop1}
\el
\ee
The latter encodes the influence of occupation and splitting of the two $\epsilon_{1,3}$ triplet levels
that have no matrix elements with the singlet ground state but contribute to the moment formation for finite
temperature below $T_m$.\\ 

For the numerical determination of the induced moment we  write, similar as for SSM and SDM $\hDe_T=\xi_tf_t(T)$.
From  Eq.~(\ref{eq:tOP}) we obtain the MF equation for $f_t(T)$ as 
\be
\bl
f_t(T)
=&
 \frac{\tanh[\bigl(\frac{\De_0}{2T}\bigr)f_t(T)]
 }{
 1+\tf_t(T)}; 
 \\
\tf_t(T)
=&
g(T)\bigl[1-\frac{2\xi_t}{m_t^2}\frac{1}{\hat{\de}_T}\tanh(\frac{\De}{2T}\hat{\de}_T)\bigr]
\label{eq:tOP2}
\el
\ee
with $\hat{\de}_T=(1/m_t)(\hDe_T^2-1)^\fs$.   The solution for $f_t(T)$ leads to $\hDe_T$ and likewise to the induced moment  $\la J_z\ra=\frac{m_t}{2}\frac{1}{\xi_t}
(\hDe_T^2-1)^\fs$ with saturation value $\la J_z\ra_0=\frac{m_t}{2}\frac{1}{\xi_t}(\xi_t^2-1)^\fs$. The latter is similar to SSM and SDM because at $T=0$ the $\Gamma_4$ contribution in the STM vanishes.\\

\begin{figure}
\includegraphics[width=0.95\linewidth]{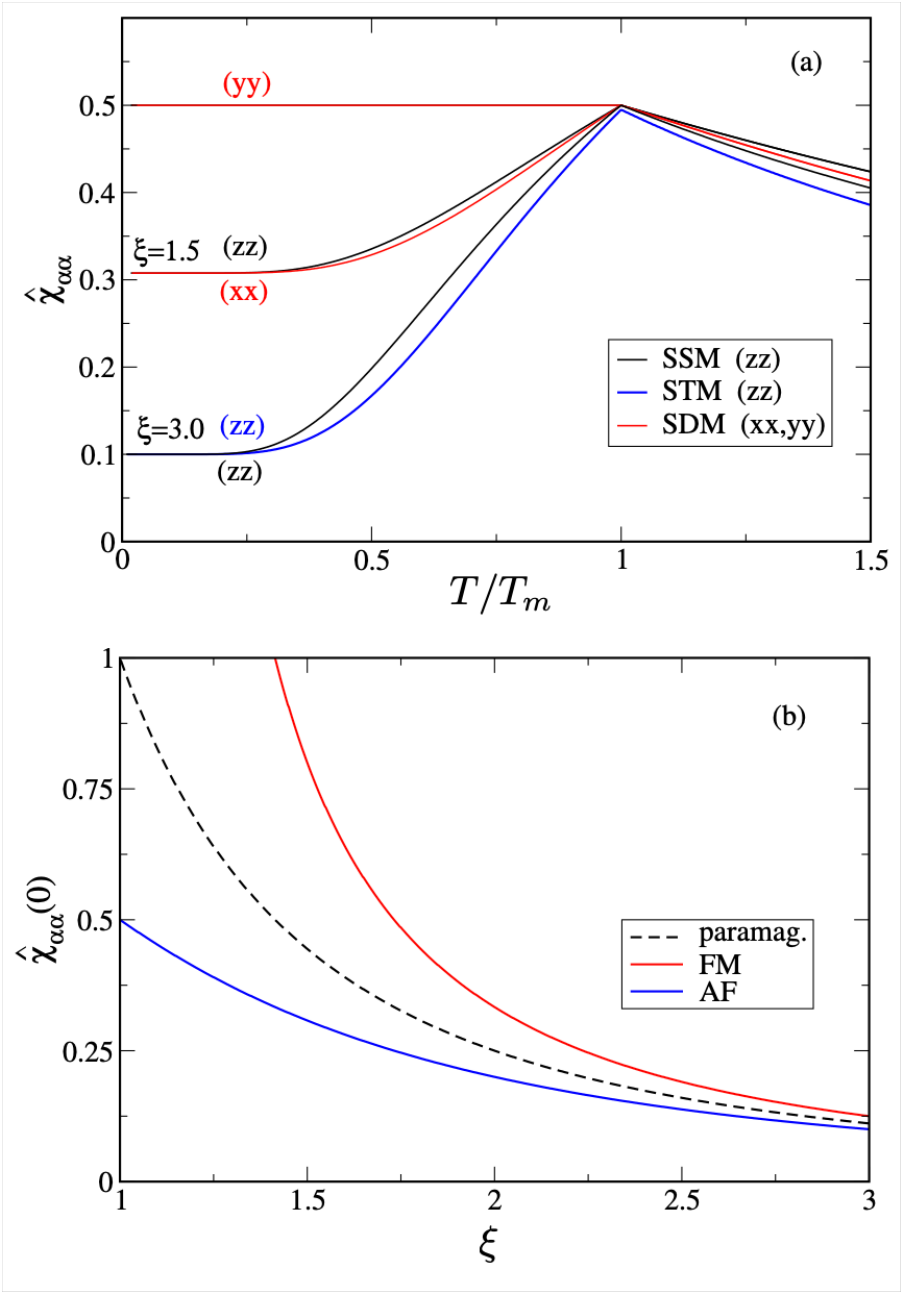}
\caption{ (a) Temperature dependence of static homogeneous susceptibilities (unit $1/I_e$) for different models and control parameters  for the AF case with $\la J_x\ra$ (SDM)  or $\la J_z\ra$ (SSM,STM) moment directions.  In the xy-type SDM the transverse (yy) susceptibility is constant below $T_m$ as in conventional degenerate ground state magnets. In contrast  the longitudinal susceptibilities at $T=0$  are generally finite, evolving from the same as transverse value at the QCP to progressively lower values with increasing $\xi$. (b) Control-parameter dependence of $T=0$ {\it longitudinal} susceptibility  as function of $\xi$, identical for the three models. It vanishes asymptotically when approaching the quasi-degenerate case.}
\label{fig:susz-T}
\end{figure}

Letting $\la J_z\ra \rightarrow 0$ in Eq.~(\ref{eq:tOP})
(note that then also $P_b\rightarrow 0$) we arrive at an implicit equation for the transition temperature of induced order given by the solution of
\be
T_m=\frac{\De}{2\tanh^{-1}(x_m)};\;\;\; 
x_m=\frac{2-\bigl(\frac{I_e}{\De}\bigr)\tanh^{-1}(x_m)}{1+\xi_t-\bigl(\frac{I_e}{\De}\bigr)\tanh^{-1}(x_m)}
.
\label{eq:tTm}
\ee
If we neglect the contribution of the  thermally excited $\epsilon_{1,3}$ levels justified for $\frac{I_e}{\Delta}=
\frac{\gamma}{m_t}=\frac{2\xi_t}{m_t^2}=0.075\xi_t \ll1$ then we arrive at the simple expression
\bea
T_m^0=\frac{\De}{2\tanh^{-1}\frac{2}{1+\xi_t}},
\label{eq:tTm0}
\eea
which is formally similar to Eqs.~(\ref{eq:sTm},\ref{eq:dTm}) of the SSM and SDM case and again $\xi_t>\xi_t^c=1$ for $T^0_m>0$.  We will show later that this
zeroth order approximation of transition temperature is identical to the temperature where the soft longitudinal (polarisation parallel to $\la J_z\ra$) exciton mode appears, However, the true transition temperature obtained from the Eq.~(\ref{eq:tTm}) is larger than $T_m^0$ due to the thermally excited Curie contributions from $|\psi_{1,3}\ra$ triplet states which stabilise the moment $\la J_z\ra$ beyond $T_m^0$. This  means that at the true $T_m>T_m^0$ the complete softening of the exciton mode will not occur but rather it will be arrested at a finite frequency (Sec.~\ref{sec:exc-STM}). This fact for the STM was known for some time~\cite{smith:72,cooper:72,buyers:75} but has not been considered in detail so far. We first calculate the approximate shift defined by $T_m=T_m^0+\delta T_m$. Using the first iteration step of Eq.~(\ref{eq:tTm}) and expanding in the small parameter $(I_e/\Delta)=2\xi_t/m_t^2$ as given above we obtain
\be
\delta T_m=\frac{1}{m_t^2}\frac{2\xi_t}{3+\xi_t}T_m^0.
\label{eq:tmshift}
\ee
This is the upward shift of the transition temperature due to the effect of diagonal matrix elements in Eq.~(\ref{eq:Jstmfield}). Their relative size is $\frac{1}{2}:\frac{m_t}{2}=\frac{1}{m_t}$ which controls the size of the correction $\delta T_m$. It is a very good approximation to the numerically determined exact shift obtained from Eq.~(\ref{eq:tTm}) shown in  Fig.~\ref{fig:Tm-xi}(a).

\subsubsection{Internal energy and specific heat for STM}

The internal energy of the STM in the paramagnetic phase is $U(T)=-\frac{\De}{2}/(2\tanh\frac{\De}{2T}-1)/
(2-\tanh\frac{\De}{2T})$ leading to a paramagnetic specific heat (corresponding to $N=3$ in Appendix \ref{app:schottky}):
\be
C_V(T)=\frac{3\bigl(\frac{\De}{2T}\bigr)^2}
{[2\cosh\bigl(\frac{\De}{2T}\bigr)-\sinh\bigl(\frac{\De}{2T}\bigr)]^2}.
\label{eq:tCV}
\ee
In the ordered phase the internal energy can be calculated as previously using a similar expression as Eq.~(\ref{eq:UT})
summing over all four states of the STM and adding the MF constant term we obtain 
\be
\bl
U(T)
&=
-\frac{\De}{2(1+g)}
\Big\{
\hDe_T\tanh
\big[
\bigl(\frac{\De}{2T}\bigr)\hDe_T
\big]
\\&
+g(T)
\Big[
\hat{\de}_T
\tanh
\big[
\bigl(\frac{\De}{2T}\bigr)\hat{\de}_T
\big]-1
\Big]
\Big\}
+\frac{\De}{4}\frac{1}{\xi_t}(\hDe_T^2-1).
\el
\ee
For zero temperature this leads to the ground state energy in the universal form
$U(0)=-\frac{\De}{4}(\xi_t+\frac{1}{\xi_t})$ and the specific heat $C_V(T)$ has to be obtained from $U(T)$ by  numerical differentiation as in the SDM case.

\subsubsection{The static homogeneous longitudinal STM susceptibility}

The static susceptibility may be calculated as before from Eq.~(\ref{eq:baresus}). 
For the induced moment phase $(T<T_m)$ we obtain $(m'_t=1)$:
\be
\bl
&
\hchi^0_{zz}(T)=
[1+g(T)]^{-1}\Bigl\{\frac{1}{I_e}\bigl(\frac{\xi_t}{\hDe_T^3}\bigr)\tanh(\frac{\De}{2T}\hDe_T) 
\\
&+
\frac{1}{T}
\Big[
\bigl(\frac{m'_t}{2}\bigr)^2 g(T)+\bigl(\frac{m_t}{2}\bigr)^2(\hDe^2_T-1)
\bigl(\frac{1}{\hDe_T^2}-\frac{1+g(T)}{\xi_t^2}\bigr)
\Big]
\Bigr\}.
\el
\ee
Note that at zero temperature with $\hDe_0=\xi_t$ and $g(0)=0$ only the first van Vleck term survives 
whereas the Curie contributions from thermally excited split triplet states vanishes.
In the paramagnetic regime $(T>T_m)$ with $\hDe_T=1$ this reduces to the explicit expression
\be
\hchi^0_{zz}(T)=\bigl(\frac{\xi_t}{I_e}\bigr)\frac{\tanh(\frac{\De}{2T})}{2-\tanh(\frac{\De}{2T})}
+\frac{1}{T}\bigl(\frac{m'_t}{2}\bigr)^2
\Bigl[
\frac{1-\tanh(\frac{\De}{2T})}{2-\tanh(\frac{\De}{2T})}
\Bigr].
\ee
Finally the homogeneous longitudinal MF-RPA susceptibility is again obtained from $ \hchi_{zz}(T) =\hchi^0_{zz}(T)/(1\mp I_e\hchi^0_{zz}(T))$ with upper and lower signs corresponding to FM or AF exchange, respectively. In the $T=0$ limit this recovers $ \hchi_{zz}(0) =(1/I_e)(\xi_t^2\mp 1)^{-1}$ which is identical to the longitudinal susceptibility in both SSM and STM because in this limit there is no contribution from depopulated excited states. In the large $\xi_t$ (quasi-degenerate) case that approaches the conventional magnet the longitudinal susceptibility vanishes because the saturation moment approaches the maximum value and no further polarisation by external field is possible (Fig.~\ref{fig:susz-T}), see also Sec.~\ref{sec:discussion}.

\section{Magnetic exciton dispersions and soft mode behaviour}
\label{sec:exc}

In the two-level SSM, SDM and STM the dynamics is described by  collective modes termed 'magnetic excitons'. They correspond to propagating CEF excitations which develop a dispersion due to effective intersite exchange. Their dispersion is strongly temperature dependent controlled by the thermal population difference of singlet ground state and excited multiplet. Within RPA approach they may exhibit, as a precursor phenomenon,  a complete softening at the magnetic ordering wave vector when $T_m$ is approached from above. Such magnetic exciton softening to varying degree has been found in quite a number of singlet ground state CEF systems, in particular Pr compounds~\cite{buyers:75,clementyev:04}. It is frequently incomplete because the static susceptibility contributions from higher lying CEF states lead to a magnetic transition already before the softening of the lowest mode is achieved. Furthermore dynamical effects beyond the RPA complicate this simple picture~\cite{jensen:91}. The magnetic exciton formation has mostly been studied in the paramagnetic phase. Here we give a complete theory for the three models also in the induced moment regime. We focus on a representation that highlights the role of the control parameters that measure the distance from the QCP and emphasise the connection to thermodynamic properties.

{\BLU Before, however, we give a simple intuitive picture (restricting to SSM) of these  paramagnetic exciton modes to distinguish them from the quasiclassical long wave length magnons which correspond to precession of ordered moments  around the moment direction. In the present case we rather have to start from a local singlet-singlet excitation $|0\rangle_i \rightarrow |1\rangle_i$ in the {\it paramagnetic} phase. Then the exchange coupling terms $J_{ij}J_z(i)J_z(j)$ between sites  ${\bf R}_i$ and  ${\bf R}_j$ allow a process where a de-excitation $|1\rangle_i \rightarrow |0\rangle_i$ at site ${\bf R}_i$ is followed by another excitation  $|0\rangle_j\rightarrow |1\rangle_j$ at the neighbouring site. Thus the singlet-singlet excitation has propagated by one lattice site due to the intersite exchange. Considering this process in translational invariant way leads directly to the dispersive
CEF modes whose energy is centerd around the local singlet-singlet excitation energy $\Delta$.}\\

The magnetic exciton modes may be calculated by the RPA dynamic response function technique~\cite{fulde:72,jensen:91,thalmeier:02} or with the bosonic Bogoliubov transformation approach~\cite{grover:65,cooper:72,liu:19,thalmeier:21,akbari:23}. Here we prefer the former because it gives a more reliable description of dispersions in the whole temperature range. The starting point is the dynamical single-ion susceptibility tensor ($\alpha,\beta=x,y$) defined by
\be
\hchi_{0\al\beta}(\om)=\sum_{n\neq m}\frac{\la m| J_\al|n\ra\la n| J_\beta|m\ra}{\epsilon_m-\epsilon_n-\om}(p_n-p_m),
\label{eq:baredynsus}
\ee
and the RPA cartesian susceptibility tensor is then obtained as
\be
\tensor{\hchi}(\bq,\om)=\bigl[1-I(\bq)\tensor{\hchi_0}(\om)\bigr]^{-1}\tensor{\hchi_0}(\om),
\label{eq:dynsus}
\ee
{\BLU In practice we will need only the diagonal components for SSM and SDM as the nondiagonal ones vanish. For the STM models there are nondiagonal transverse $xy$ components in the ordered phase with induced moment $\la J_z\ra$. However we will consider only the longitudinal modes in this case which are obtained from the diagonal response function $\hchi_{zz}$ only.}
The poles of these response functions determine the temperature-dependent magnetic exciton modes. This formulation is valid both in the paramagnetic and magnetic regimes provided the proper split level energies and matrix elements for the three models are used. In the magnetic phase we will restrict to FM case $(I_e>0)$ because for two - sublattice AFM order the dimension of the susceptibility matrix is doubled to four, making it more involved for analytical treatment. The AFM case has been considered before for the Ising- type SSM~\cite{thalmeier:02}.  

{\BLU The only method capable of investigating magnetic exciton dispersions experimentally is inelastic neutron scattering (INS) where the scattering cross section is proportional to the imaginary part of the above
response functions and exhibits peaks at its poles that depend on momentum transfer which allow determination of the mode dispersion.
Since the energies involved may be quite low in the few meV range (in particular for the most interesting incipient soft modes) the more recent resonant x-ray scattering techniques used for investigation of high energy magnons have not yet been applied in the present context.}
 %
\begin{figure}
\includegraphics[width=0.99\linewidth]{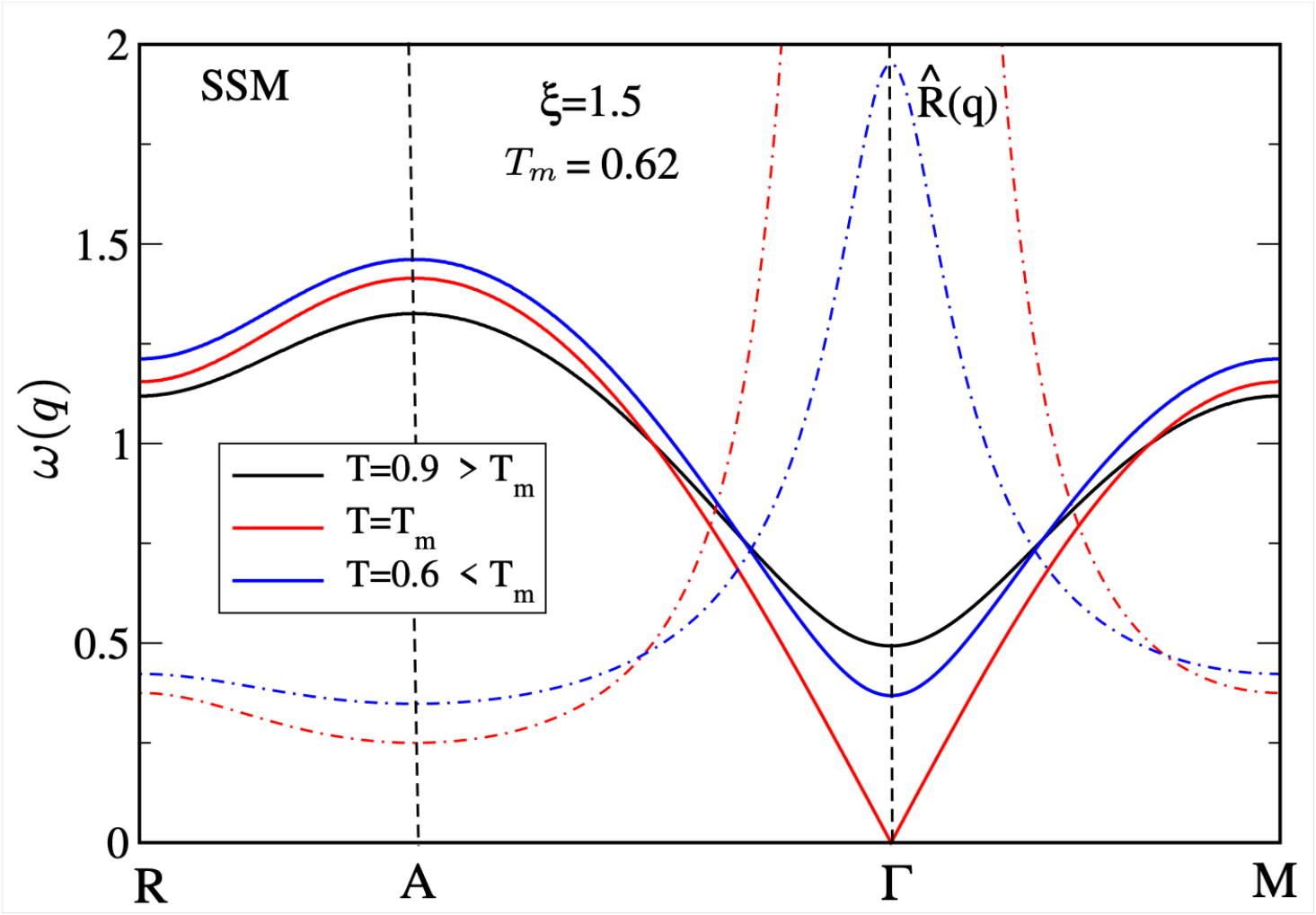} 
\caption{ (a) Magnetic exciton modes for FM case along simple tetragonal BZ path R(001), A(111), $\Gamma$(000), M(110) for SSM at three different temperatures.
A softening at the FM $\Ga$ -point occurs at $T_m$ with a subsequent  re-hardening below $T_m$. The normalised intensity $\hat{R}(\bq)$ is represented by the dash-dotted lines, showing the diverging intensity (in units $(\De/I_e)$) of the soft mode. {\BLU Here and in the following dispersion plots $\omega(\bq)$ is given in units of $\Delta$ and we use $\kappa=1$.} }
\label{fig:disps-q}
\end{figure}

\subsection{Magnetic excitons in the Ising- type  SSM}
\label{sec:exc-SSM}

Firstly we consider the SSM case,  the calculation of the zz component of the single ion dynamical susceptibility according to Eq.~(\ref{eq:baredynsus})  leads to
\be
\bl
\hchi^0_{zz}(\om)
&=
\bigl(\frac{m_s}{2}\bigr)^2\cos^2 2\theta
\tanh
\bigl(
\frac{\De_T}{2T}
\bigr)
\frac{2\De_T}{\De_T^2-\om^2},
\label{eq:susmat}
\el
\ee
where $\De_T=\De\xi_sf_s(T)=\De\hDe_T$ is the renormalised CEF splitting and $\cos2\theta=1/[\xi_sf_s(T)]$.
As described in Sec.~\ref{sec:model} for simplicity we use the intersite exchange $\hat{I}(\bq)$ for the simple tetragonal lattice with lattice constants $a,c$ set to unity and the  corresponding real-space exchange anisotropy defined by $\kappa$ (Eq.~(\ref{eq:exfunc})). However the theory may be applied to any other uniaxial  Bravais lattice by using the corresponding exchange function $\hat{I}(\bq)$.
With the above expression for $\hchi^0_{zz}(\om)$ the poles  of the collective susceptibility $\hchi_{zz}(\om)$, i.e. the magnetic exciton dispersion in the induced moment phase  may be found as
\be
\bl
\omega(\bq,T)
\!
=
&
\De_T
\Big[
1-\frac{1}{\xi_s^2f_s(T)^2}\hat{I}(\bq)\Big]^\fs
.
\label{eq:sexcgen}
\el
\ee
For $T>T_m$ the singlet mixing angle $\theta$ vanishes and we obtain
the  well known SSM paramagnetic exciton dispersion
\be
\bl
\omega(\bq,T)
&=\De\bigl[1-\xi_sf_s^0(T)\hat{I}(\bq)\bigr]^\fs ,
\label{eq:sexcpara}
\el
\ee
where  $\hat{I}(\bq)=I(\bq)/I(0)$ or  $\hat{I}(\bq)=I(\bq)/I(\bq_m)$ is the exchange Fourier transform normalised to the
maximum value  for FM $(\bq_m =0)$  or AFM $(\bq_m =(111))$ wave vectors, respectively. We note that the SSM exciton mode at the FM or AFM wave vector for $I_0>0$ or $I_0<0$ decreases continuously on approaching $T_m$ as a precursor of the ordering  and becomes soft at the transition (Fig.~\ref{fig:disps-q}) according to $\omega(\bq,T_m)=\De[1-\hat{I}(\bq)]^\fs$. 
On the other hand at zero temperature the mode dispersion in Eq.~(\ref{eq:sexcgen}) reduces to the simple form $(\De_0=\xi_s\De)$
\be
\omega(\bq,0)=
\De_0\bigl[
1-\frac{1}{\xi_s^2}\hat{I}(\bq)]^\fs.
\ee
Therefore at the ordering wave vector with $\hat{I}(\bq_m)=1$  the paramagnetic exciton  mode becomes soft at $T_m$ and re-hardens below it since $\xi_s>1$. The INS intensity $R(\bq,T)$ of a dispersive exciton mode is not constant but varies with momentum and temperature (without the Bose prefactor) according to
\be
\bl
R(\bq,,\omega,T)
&=
 \frac{1}{\pi}
 {\rm Im }
 \hchi_{zz}(\bq,\om)=
 \hat{R}(\bq,T)\delta(\omega-\omega_\bq),
\el
\ee
with the weight function given in the ordered phase by $(f_s=f_s(T))$:
\be
\bl
\hat{R}(\bq,T<T_m)=
\frac{\De^2}{2I_e\omega(\bq,T)}=
\fs\bigl(\frac{\De}{I_e}\bigr)
[\xi_sf_s-\hat{I}(\bq)]^{-\fs}
.
\label{eq:sintens1}
\el
\ee
and in the paramagnetic phase we obtain
\be
\bl
\hat{R}(\bq,T>T_m)=
\fs\bigl(\frac{\De}{I_e}\bigr)
(\xi_sf_s^0)
[1-(\xi_sf_s^0)\hat{I}(\bq)]^{-\fs}
.
\label{eq:sintens2}
\el
\ee
 At the transition temperature with $\xi_sf_s=\xi_sf_s^0=1$ then  $\hat{R}(\bq,T_m)=\fs\bigl(\frac{\De}{I_e}\bigl)[1-\hat{I}(\bq)]^{-\fs}$ which diverges at the soft mode wave vector ($\Gamma$ point). The behaviour of $\hat{R}(\bq,T)$ in the magnetic phase showing the singularity is plotted in Fig.~\ref{fig:disps-q} with a broken line.
%
%
\begin{figure}
\includegraphics[width=0.99\linewidth]{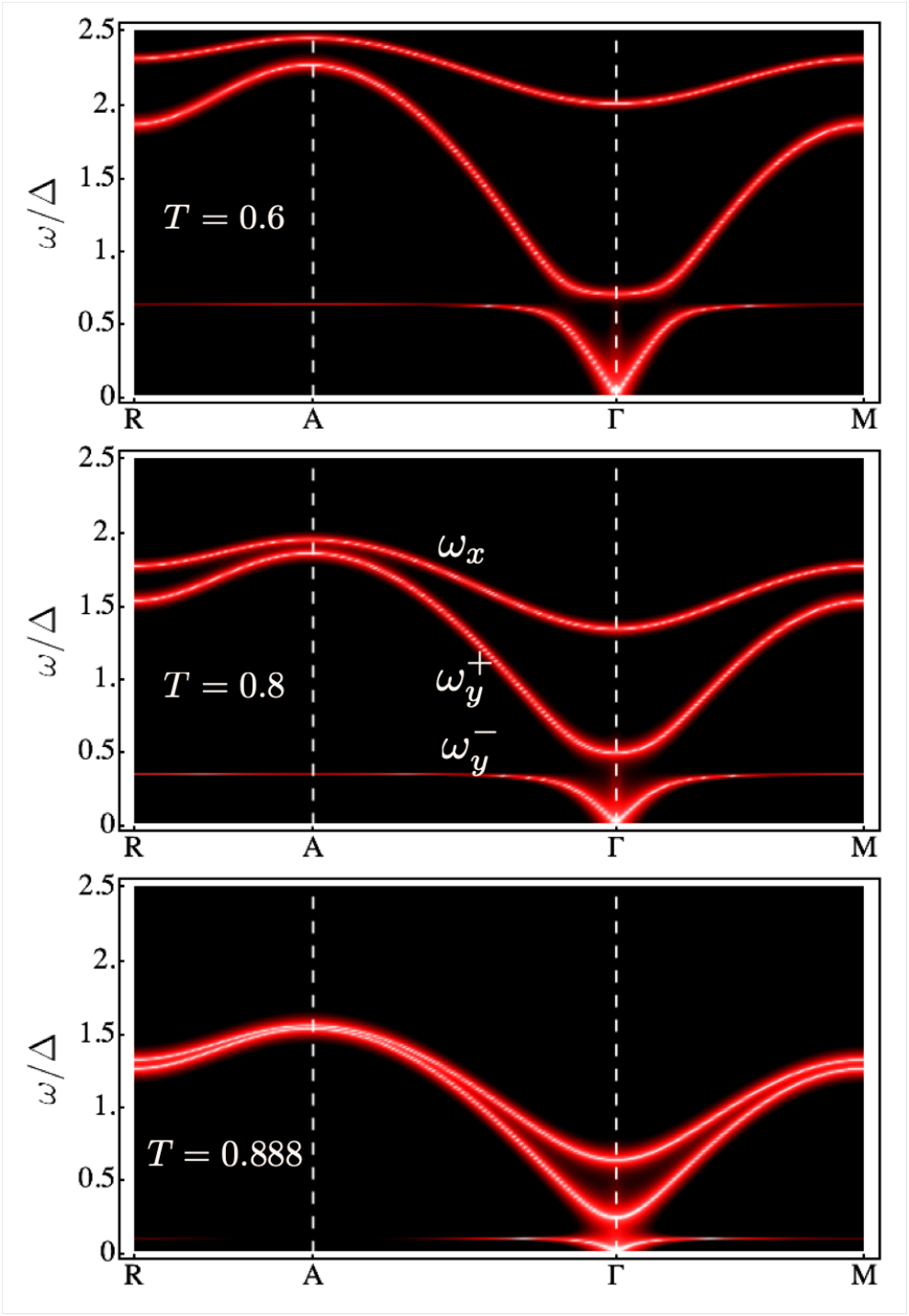} 
\caption{SDM spectral density (Eq.~(\ref{eq:dspecdens})) of magnetic exciton branches in the induced moment phase with $\xi_d=2.5$ for different temperatures $T<T_m=0.91$. To enhance visibility $\ln R(\bq,\omega)$ is plotted and a broadening of $\eta =0.005$ is used. Three modes appear due to the excited doublet state. The high-energy ones are due to singlet-doublet excitations while the lower one corresponds to thermally excited transitions between the doublet components. It exhibits hybridisation and anti-crossing with one of the high-energy modes and its intensity is appreciable only in that region (see also Fig.~\ref{fig:dhyb-T}(a)). The soft mode actually originates from the anti-crossing high energy part. For $T\rightarrow T^-_m$ the hybridisation gap is closed and $\omega^-_y(\bq)$ vanishes for all wave vectors due to the degeneracy of the excited doublet components  (see also Fig.~\ref{fig:dhyb-T}(b)). For $T>T_m$ (not shown) only one fully gapped (degenerate $\omega_x,\omega_y$) branch remains.}
\label{fig:dispd-q}
\end{figure}

\subsection{Magnetic excitons in the xy-type SDM}
\label{sec:exc-SDM}

The calculation of exciton dispersion in the induced moment state of this extended model  is more involved. Firstly there is also a CEF excitation from the thermally populated $\epsilon_2$ level that has to be taken into account in principle, although it becomes insignificant at low temperatures. Furthermore for $T<T_m$ two excitation branches from the singlet ground state with different energies occur. On the  other hand $J_x$, $J_y$ operators in the eigenvector basis of the induced moment phase (Eq.~(\ref{eq:Jsdm})) have matrix elements for mutually exclusive transitions so that again no mixed nondiagonal dynamic susceptibilities occur. The diagonal elements $\hchi^0_{\al\al}$ are given by 
\be
\bl
\hchi^0_{xx}(\om)
&=
\frac{m_d^2m_1^2\De_Tf_d(T)}{\De_T^2-(\om)^2},
\\
\hchi^0_{yy}(\om)
&=
\frac{m_d^2m_2^2\De_Tf^{12}_d(T)}{(\De^{21}_T)^2-(\om)^2}+
\frac{m_d^2m_2^{'2}\De^{32}_Tf^{23}_d(T)}{(\De^{32}_T)^2-(\om)^2},
\el
\ee
where $\Delta_T=\epsilon_3-\epsilon_1=(\xi_d\Delta)f_d$ and the matrix elements $m_1, m_2, m_2'$ are given in Eq.~(\ref{eq:Jsdm2}). The additional transition energies 
and their associated occupation differences for the $yy$ case are defined by $\Delta_T^{21}=\epsilon_2-\epsilon_1=\fs(\De_T+\De)$, $\Delta_T^{32}=\epsilon_3-\epsilon_2=\fs(\De_T-\De)$ and similar $f_d^{12}(T)=p_1-p_2$, $f_d^{23}(T)=p_2-p_3$. They fulfil the constraints $\De_T^{21}+\De_T^{32}=\De_T$ and $f_d^{12}+f_d^{23}=p_1-p_3=f_d(T)$. The diagonal RPA susceptibility matrix elements are then simply given by $(\al=x,y)$
\be
\hchi_{\al\al}(\bq,\om)=\frac{\hchi^0_{\al\al}(\bq,\om)}{1-I(\bq)\hchi^0_{\al\al}(\bq,\om)}
\ee
Their poles lead to three magnetic exciton branches, one for $xx$ and two for $yy$ polarisation. We obtain
\be
\bl
 xx:\;\;\; \omega_x(\bq,T)=
\left\{
\begin{array}{l l}
\De_T[1-\frac{1}{(\xi_df_d)^{2}}\hat{I}(\bq)]^\fs;\;\;\;T<T_m\\[0.3cm]
\De[1-\xi_df_d\hat{I}(\bq)]^\fs;\;\;\;T>T_m
\end{array}
\right.
.
\label{eq:xxmode0}
\el
\ee
%
\begin{figure}
\includegraphics[width=0.90\linewidth]{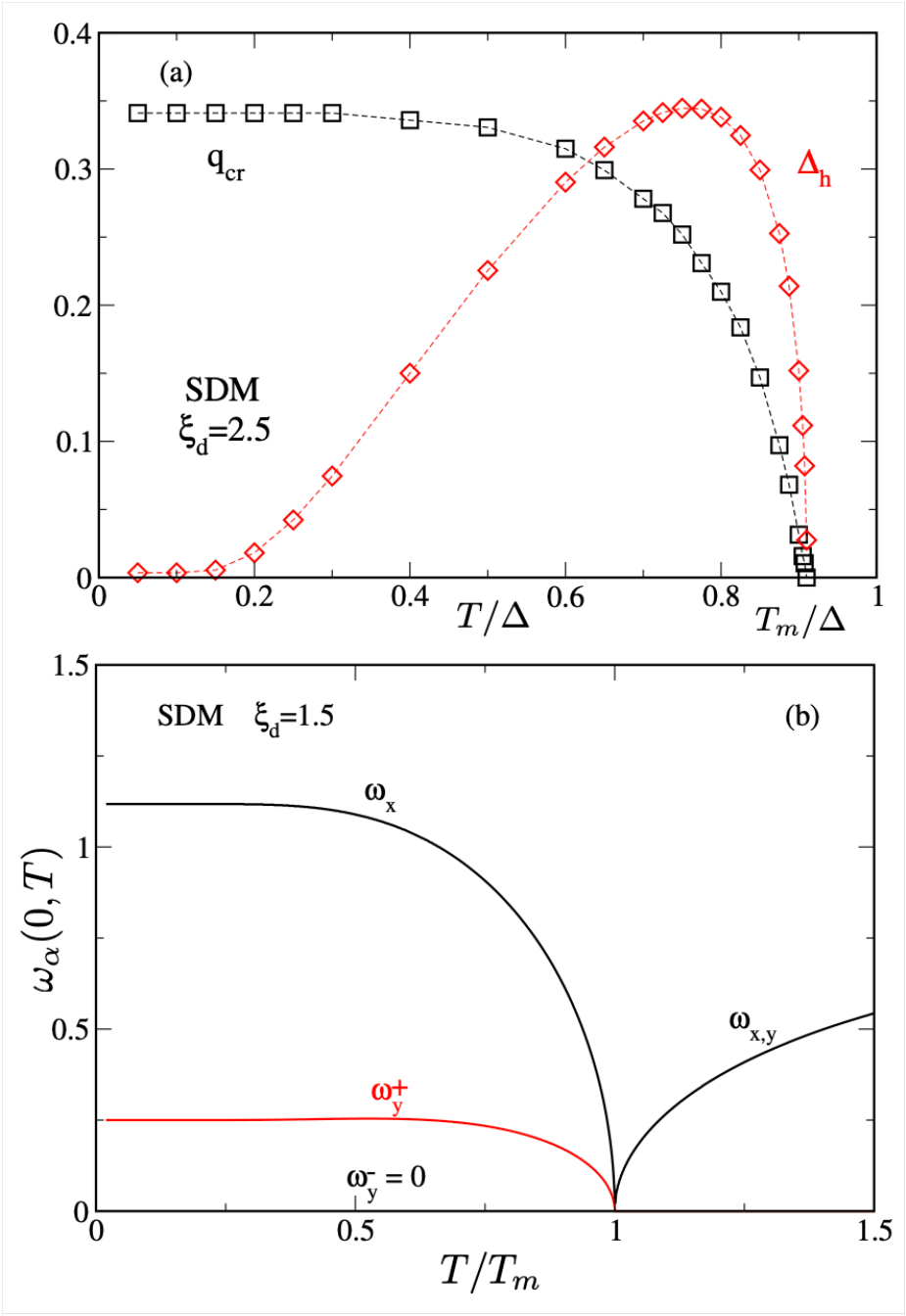} 
\caption{
(a) Hybridisation behaviour between transverse $\omega_y^\pm(\bq)$ modes as function of temperature in the induced moment phase. Here $q_{cr}=|\bq_{cr}|$ (in units of $\pi, (a,c=1)$) is the wave vector of mode crossing along (111) with respect to $\Gamma$ point and $\De_h$ the hybridisation gap that opens at the crossing point, see also Fig.~\ref{fig:dispd-q}. The hybridisation gap is non-monotonic and achieves its maximum for an intermediate temperature. It vanishes for low T due to thermal depopulation of doublets and for $T_m$ where the modes themselves become soft and their gap also has to close. Dashed lines are guides to the eye.
(b) Soft mode $(\bq=0)$ evolution with temperature from paramagnetic to ordered state $(\xi_d=1.5)$. Two of the modes re-harden below $T_m=0.51$ and the remaining one stays at zero energy.}
\label{fig:dhyb-T}
\end{figure}
This result  corresponds formally to the Ising-type SSM (Eqs.~(\ref{eq:sexcgen},\ref{eq:sexcpara})). because the $J_x$ dipolar operator in Eq.~(\ref{eq:Jsdm}) connects only to the $\epsilon_3$ level (which furthermore has no matrix element for $J_y$ so that no mixed response function appears). Therefore it is formally equivalent to an Ising type singlet singlet model (for $J_x$), the only instance where the doublet nature of the excited state enters is through the population difference factor $f_d(T)$, but not in the dynamics. 

On the other hand for the $yy$ response function two more modes appear originating from ground state to $\epsilon_2$ excited level and from this to the upper level $\epsilon_3$. The latter is a thermally activated mode which looses its intensity for
low temperature. We obtain:
\be
\bl
yy:\;\;\; \omega^{y2}_\pm(\bq,T) 
&=
\fs B(\bq,T)\pm[\frac{1}{4}B(\bq,T)^2-C(\bq,T)]^\fs;
\\
C(\bq,T)
&=\bigl(\frac{\De}{2}\bigr)^4(\hDe_T^2-1)^2[1-\hat{I}(\bq)];
\\
B(\bq,T)
&=
\bigl(\frac{\De}{2}\bigr)^2\Bigl\{2(1+\hDe_T^2)
\\
&
\hspace{-1cm}
-
\bigl(\frac{\xi_d}{\hDe_T}\bigr)
\bigl[(\hDe_T+1)^2f_d^{12}+(\hDe_T-1)^2f_d^{23}\bigr]\hat{I}(\bq)\Bigr\}
,
\label{eq:yymode0}
\el
\ee
where $\pm$ correspond to the high energy mode and thermally excited low energy mode with 
maximum or vanishing intensity at zero temperature, respectively. {\BLU The auxiliary quantities B,C are the
coefficients for the linear and constant term in the quadratic equation for the exciton frequency}. \\

We now discuss a few important special cases. At $T=T_m$ with $\hDe=1$ this reduces to $C(\bq,T)=0$ and $f_d^{23}=0,f^{12}_d=f_D=1/\xi_d$. Then we simply get
\be
 \omega_\pm^y(\bq,T=T_m)= 
\left\{
\begin{array}{l l}
\De[1-\hat{I}(\bq)]^\fs \equiv \omega_x(\bq,T_m)\\[0.3cm]
0
\end{array}
\right.
.
\ee
 Therefore in the paramagnetic case the $x$ and $y+$ modes are degenerate and become soft at $T_m$ (Fig.~\ref{fig:dhyb-T}(b)). The $y-$ mode corresponding to the transitions between degenerate doublet states is a thermally activated zero energy mode for all \bq~ (quasielastic mode in reality). At zero temperature for saturated  order $(\hDe_T=\xi_d, f_d=f_d^{12}=1, f_d^{23}=0)$ the doublet is split and both $y$-modes have nonzero energy given by
\be
 \omega^y_\pm(\bq,T=0)= 
\left\{
\begin{array}{l l}
\frac{\De}{2}(\xi_d-1)\\[0.3cm]
\frac{\De}{2}(\xi_d+1)[1-\hat{I}(\bq)]^\fs
\end{array}
\right.
.
\label{eq:dsoft0}
\ee
This holds for $|\bq|<|\bq_{cr}|$, for larger $|\bq|$ the $\pm$ mode labels are interchanged. Here $|\bq_{cr}|=\sqrt{3}\cos^{-1}(4\xi_d/(\xi_d+1)^2)$ is the wave vector where the two modes are crossing for $T=0$ along $(111)$ or $\Ga A$ direction and hybridising for
finite temperature (see Fig.~\ref{fig:dispd-q}(a)). At the ordering wave vector $\bq_m=0$ Eq.~(\ref{eq:dsoft0}) reduces to  $\omega^y_-(0,T=0)= 0$ and $\omega^y_+(0,T=0)= \frac{\De}{2}(\xi_d-1)$.
The above equations demonstrate that $\omega^y_-(\bq=0,T)$ vanishes at both $T=0,T_m$, in fact it is a Goldstone soft mode within the whole ordered regime $T\leq T_m$ since $\hat{I}(\bq_m)=1$ in Eq.~(\ref{eq:yymode0}). This is due to the fact that $(J_x,J_y)$ moments have continuous rotation symmetry in the SDM, whereas in the Ising-type SSM without such continuous symmetry the soft mode re-hardens immediately below $T_m$. The second transverse mode, however shows stiffening for $T<T_m$ described by
\be
\bl
&\omega_+(\bq=0,T)
\!=
\\
&\;\;
\frac{\De}{2}\Bigl\{2(1+\hDe_T^2)
-\!
\frac{\xi_d}{\hDe_T}
\bigl[
4\hDe_Tf_d^{12}(T)
\!+\!
(\hDe_T-1)^2f_d(T)]\Bigl\}^\fs
.
\el
\ee
This frequency starts at zero for $T_m$ and below increases to the maximum value $\frac{\De}{2}(\xi_d-1)$ at $T=0$  (Eq.~(\ref{eq:dsoft0})). The dispersion and temperature behaviour  of the exciton modes in the SDM model are presented in Figs.~\ref{fig:dispd-q} as spectral plots and will be further discussed in the following section.

\begin{figure}
\includegraphics[width=0.940\linewidth]{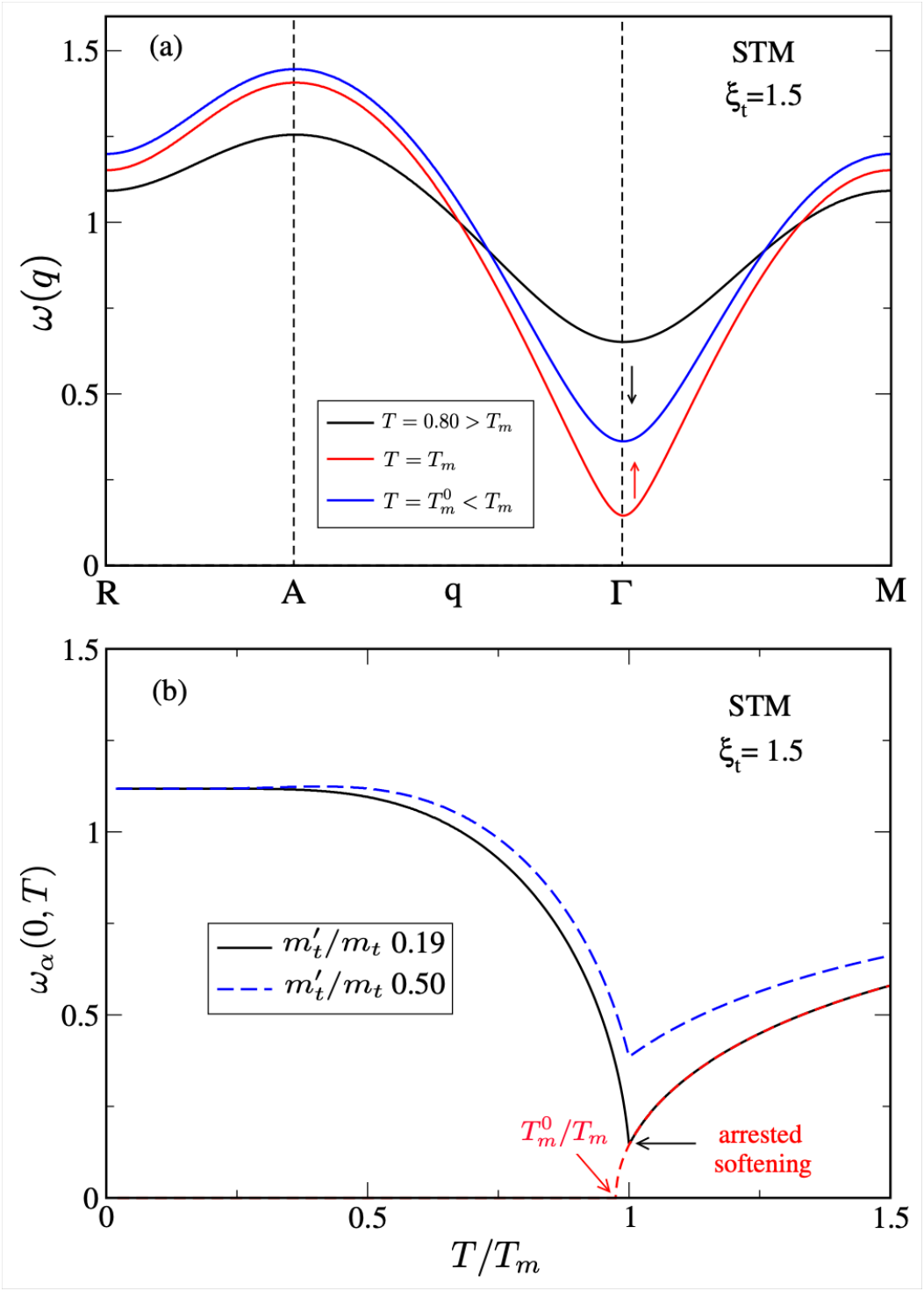} 
\caption{(a) STM exciton dispersion of longitudinal mode, At the $\Gamma-$ point first a softening (black arrow)  is observed but is arrested at $T_m$ at finite energy; followed by a re-hardening below $T_m$ (red arrow).
(b) Arrested soft mode ($\bq=0$) temperature evolution. The extrapolated paramagnetic mode touches zero at $T^0_m/T_m=0.975$ $(T_m=0.467)$ where $T_m^0$ is the approximate ordering temperature in  (Eq.~(\ref{eq:tTm0})) by neglecting the excited triplet Curie terms.}
\label{fig:dispt-q}
\end{figure}

\subsection{Magnetic excitons in the cubic STM}
\label{sec:exc-STM}

The dynamical longitudinal single ion susceptibility $\chi^0_{zz}(\bq,\om)$ has only contributions from the transition between
$\epsilon_\pm$ states but no others, even in the magnetic phase as can be seen from the $J_z$ matrix in Eq.~(\ref{eq:Jstmfield}). Therefore it is similar in structure to  the SSM case with 
\be
\chi^0_{zz}(\bq,\om)=\frac{\xi_t}{I_e}\frac{1}{\hDe_T}\frac{\De^2}{(\De_T^2-\om^2)}P_a(T)
.
\ee
Using this expression and inserting into Eq.~(\ref{eq:dynsus}) the resulting pole give the longitudinal exciton mode
dispersions $(\hDe_T=\xi_tf_t(T))$:
\be
\bl
\omega(\bq,T)=
\left\{
\begin{array}{r l}
\De_T[1-\frac{\xi_t}{\hDe^3_T}P_a(T)\hat{I}(\bq)]^\fs;\;\;\; T<T_m\\
\De[1-\xi_tP^0_a(T)\hat{I}(\bq)]^\fs;\;\;\; T>T_m
\end{array}
\right.
.
\label{eq:tdisp}
\el
\ee
Here again $\hat{I}(\bq)=I(\bq)/I_e$ is the exchange normalised to that at the ordering vector $\bq_m$ for FM $(\bq_m=0)$ or AFM $(\bq_m=(\pi,\pi,\pi)$ for which then  $\hat{I}(\bq_m)=1$. Using the lower paramagnetic expression and $P_a^0(T)$ from Eq.~(\ref{eq:tpop1}) we see that at the ordering vector the mode energy becomes soft  with $\omega(\bq_m,T_m^0)=0$ at the approximate transition
temperature $T_m^0$ of Eq.~(\ref{eq:tTm0}) which is lower than the real $T_m$ obtained from Eqs.~(\ref{eq:tTm}). In reverse this means that the mode energy will be arrested at a finite value at $T_m>T_m^0$ and the softening is incomplete. Using the first iterative approximation for $T_m$ from Eq.~(\ref{eq:tTm}) this finite value is given by approximately by
\be
\bl
&
\omega^2(\bq_m,T_m)=
\bigl(\frac{\De^2}{2m_t^2}\bigr)
(\xi_t^2-1)\tanh^{-1}\bigl(\frac{2}{1+\xi_t}\bigr)
\\
&
\rightarrow 
\omega(\bq_m,T_m)\simeq\bigl(\frac{\De}{m_t}\bigr)\delta^\fs|\ln\frac{\delta}{2}|^\fs
\label{eq:softapprox}
\el
\ee
where the second expression is the asymptotic form close to the QCP $(\xi_t=1+\delta;\;\delta\ll 1)$. Therefore $\omega(\bq_m,T_m)$ is directly proportional to the ratio $m'_t:m_t =1/m_t$ of diagonal (responsible for the shift of $T_m$) and non-diagonal (responsible for induced order) matrix elements in the $J_z$ matrix of Eq.~(\ref{eq:Jstm}). In the $J=4$ STM used here it is $1/m_t=0.145 \ll 1$ and therefore $\omega(\bq,T_m)\ll\De$ for a reasonably sized $\xi_t$. But for a general $\Gamma_1-\Gamma_4$ reduced level scheme $m'_t/m_t$ depends on the CEF potential parameters and may vary. Below $T_m$ the mode energy increases sharply again due to the effect of the molecular field.  The dependence of the arrested soft mode frequency on this ratio of diagonal to non-diagonal matrix elements and on the control parameter $\xi_t$  is shown in the two panels of Fig.~\ref{fig:softmode-t}, respectively. We note that in the ordered phase the components $J_x,J_y$ transverse to the chosen moment direction $\la J_z\ra$ have transitions between the excited triplet states. Therefore, as in the SDM case the transverse exciton modes for the STM would consist of several branches hybridising with each other, similar as in Fig.~\ref{fig:dispd-q}.

\section{Discussion of numerical results}
\label{sec:discussion}

In the following  discussion we will focus on the most typical results for the three models that show the characteristic distinction of induced moment magnetism as compared to common quasi-classical magnetic order. Therefore we will not  present and discuss figures for all physical quantities for all three models that can be obtained from the previous analysis.\\

As a prerequisite we depict the CEF level splittings in Fig.~\ref{fig:levels-T} for the three singlet ground state models caused by the appearance of the T- dependent induced order parameter $\la J_x\ra_T$ or $\la J_z\ra_T$ below $T_m$.  For SSM a symmetric repulsion of the two singlets due to their mixing  is observed such that the splitting increases to $\De_0=\xi_s\De$ at $T=0$. For the SDM only the symmetric combination of the excited doublet states mix with the ground state singlet and show the similar repulsion to an increased $T=0$ splitting of $\De_0=\xi_d\De$. The second antisymmetric doublet combination remains isolated at the paramagnetic $\De$, nevertheless it influences the thermodynamic properties by its thermal population. In the STM the repelling singlet and triplet component behave similar as in STM again with a $\De_0=\xi_t\De$. The other two states do not mix with the ground state singlet but split due to their diagonal matrix element $(m'_t/2)$ with a splitting energy directly proportional to the order parameter $\la J_z\ra$, therefore it starts with an infinite slope at $T_m$ while that of the singlet ground state and upper triplet component begin with a finite slope.\\
\begin{figure}
\includegraphics[width=\linewidth]{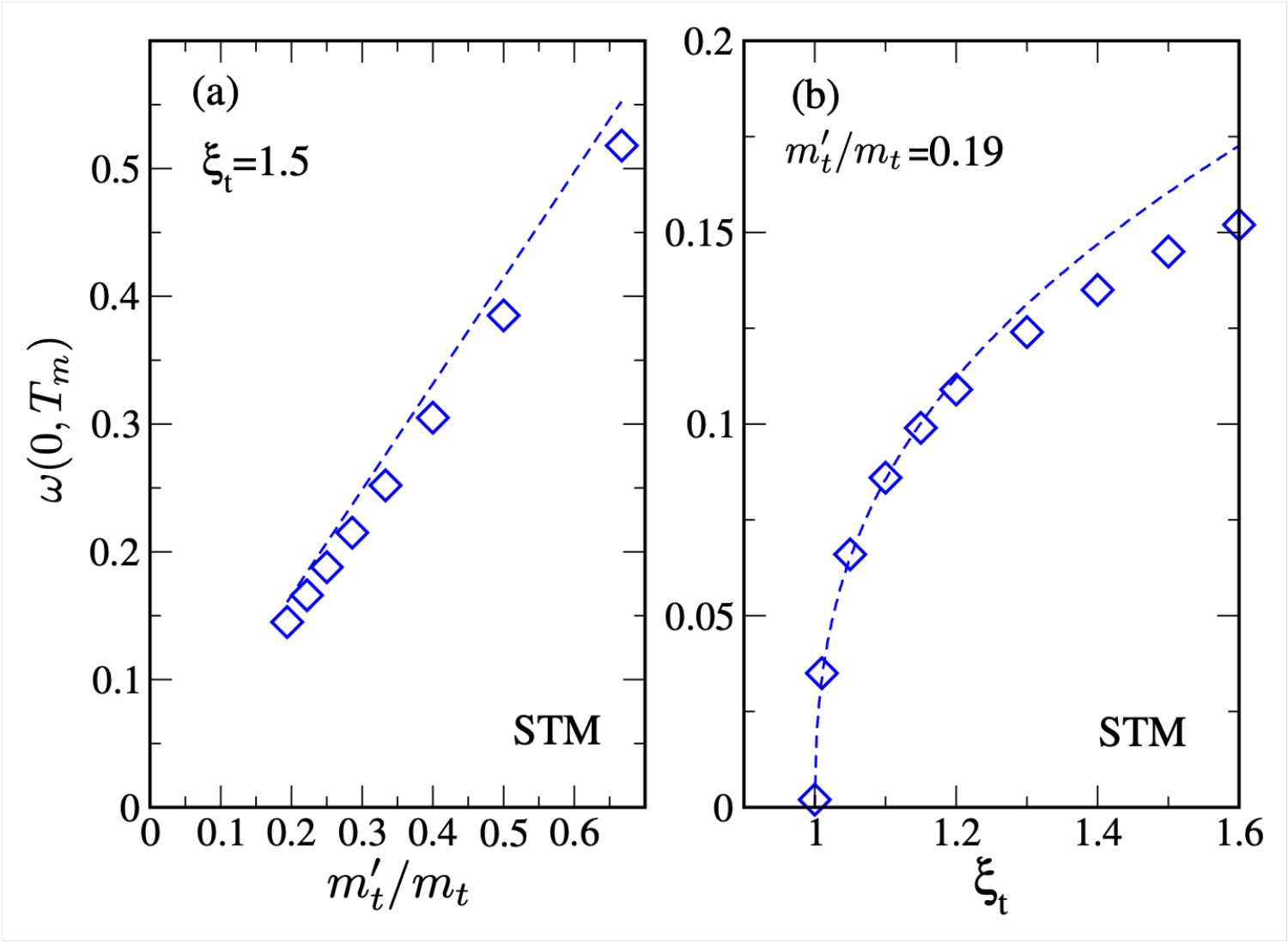} 
\caption{Arrested soft mode frequency as function of ratio of diagonal $\Gamma_4$ $(m'_t=1)$  to nondiagonal 
$\Gamma_1-\Gamma_4$ $(m_t)$ matrix elements (a) for $\xi_t=1.5$  and as function of control parameter (b) for J=4 ratio $m'_t/m_t=0.19$. Diamonds: exact numerical results from Eq.~(\ref{eq:tdisp}). Dashed lines: approximate result from Eq.~(\ref{eq:softapprox}).}
\label{fig:softmode-t}
\end{figure}
A central characteristic of induced moment magnetism is the fact that to achieve a finite transition temperature the control parameter must exceed the critical value $\xi_c=1$. This defines the QCP separating paramagnetism $(\xi<\xi_c)$ from induced quantum magnetism $(\xi>\xi_c)$. The dependence of $T_m$ on the control parameter is shown in Fig.~\ref{fig:Tm-xi}(a) for the three models as obtained from Eqs.~(\ref{eq:sTm},\ref{eq:dTm},\ref{eq:tTm}). For STM we show the exact numerical solution of $T_m$ from solving Eq.~(\ref{eq:tTm}) (full line) as well as the zeroth order approximation $T_m^0$ (broken line) given by Eq.~(\ref{eq:tTm0}). Their difference is quite small because the ratio of elastic (triplet) to inelastic (singlet-triplet) matrix elements is  $m'_t/m_t =0.19 \ll1$ and the difference is well described by the first iterative improvement in Eq.~(\ref{eq:tTm})  leading to Eq.~(\ref{eq:tmshift}). Despite its smallness it plays an essential role in the arrested soft-mode behaviour of the STM exciton modes as discussed below. For $\xi\rightarrow 1^+$ there is a logarithmic singularity of the slope of $T_m(\xi)$ and for large $\xi$ the dependence on $\xi$ becomes linear (see Eq.~(\ref{eq:sTm})). The logarithmic singularity has a practical consequence for identifying induced moment magnetism. It shows that for achieving $T_m < 0.2\De$ one must fine tune $\xi$ very close to the QCP. Therefore if this inequality is observed it is a priori unlikely that the mechanism of induced order can be invoked for a given compound. Such interpretation holds also for nonmagnetic (multipolar) induced order. In this case one should consider other mechanisms, e.g. based on hybridisation and itineracy of f-electrons for the observed order.\\

For the same control parameter $\xi$ the $T_m$ of the three models are quantitatively different. However, if we scale temperature with each individual $T_m$ it turns out that the temperature dependence of their order parameters is quite similar as shown in Fig.~\ref{fig:Tm-xi}(b). The panel (b) depicts the corresponding  functions $f_{s,d,t}(T)$  which are the difference of thermal populations between the states connected by the nondiagonal $m_{s,d,t}$ matrix elements.These functions depend on $\xi_{s,d,t}$ and are central for the calculation of most physical quantities. They are obtained from solving numerically their selfconsistency equations.  Below $T_m$ they may also be interpreted as the T-dependent normalised energy splitting $\hDe_T$ for the $m_{s,d,t}$ - connected levels below $T_m$.\\

A most characteristic feature of  induced moment magnetism is the behaviour of the specific heat $C_V(T)$  and in particular its jump at $T_m$. It is presented in Fig.~\ref{fig:CVT}(a) for SSM for three different control parameters together with the internal energy $U(T)$ and in complementary Fig.~\ref{fig:CVT}(b) for the three models with the same $\xi$. The contribution from the order parameter T-derivative that leads to the jump  (Eq.~(\ref{eq:sCV1})) is seen to be superimposed on the Schottky type background specific heat $C_V^0$ (Eqs.~(\ref{eq:sCV},\ref{eq:dCV},\ref{eq:tCV})). This means on lowering temperature the entropy is released by the induced moment transition as well as the single ion depopulation effect of the excited CEF level. The jumps increase with the degeneracy of the excited CEF level while the corresponding $T_m$ decreases (Fig.~\ref{fig:CVT}(b)).

On approaching the QCP from  above while $T_m$ shifts to lower value the specific heat jump moves with it and becomes progressively smaller. This dependence is also seen directly in Fig.~\ref{fig:dCV-xi} as continuous drop in $\delta C_V$  close to the QCP with $\xi=1+\de$ ($\de\ll 1$).  While the moment behaves like $\la S_\al\ra\sim (\de/2)^\fs$ (Eq.~(\ref{eq:momasym})) with a singular slope at the QCP the specific heat jump $\de C_V\sim (\de^2/2)|\ln(\de/2)|^3$ (Eq.~(\ref{eq:sCV1})) approaches zero more gradually. The similar though quantitatively different behaviour for SDM and STM from numerical calculations is also shown.
The $\xi$- dependent reduction of  $\delta C_V(T_m)$ in accordance with  the normalised ordered saturation moment $\la S_\al\ra_0$ is distinct from quasi-classical magnets where the saturation moment and specific heat jump are just constants independent of interaction parameters. On the other hand for large $\xi$ one moves to this quasi-classical regime  with constant $\delta C_V=\frac{3}{2}$ (for effective $S=\frac{1}{2}$ for SSM). Finally we note that not only the individual jump $\delta C_V(T_m)$ and  $C_V^+(T_m)$ approach zero for $\xi\rightarrow 1^+$ their ratio also tends to zero according to
$\de C_V/C_V^+\sim\de|\ln\frac{\de}{2}|$.\\

A similar important feature of distinction to quasi-classical  AF magnets is seen in Fig.~\ref{fig:susz-T}. In the SDM below $T_m$ the transverse (with respect to moment direction) susceptibility stays constant while the longitudinal one falls to zero at low temperature where the saturated maximum moment can no longer be polarised. For the singlet ground state quantum magnet the transverse behaviour is the same while the longitudinal one behaves fundamentally different. When we start with $\xi$ close above the QCP $\xi_c=1$ the latter is only slightly less than the transverse one because the induced saturation moment $\la S_x\ra_0$ (Fig.~\ref{fig:CVT}(b)) is very small and therefore may be easily polarised by the probe field leading to a similar large longitudinal susceptibility. However, as seen in Fig.~\ref{fig:CVT}(b) when the control parameter $\xi$ is increased the saturation moment increases rapidly and hence the low temperature susceptibility drops to lower values  as demonstrated in  Fig.~\ref{fig:susz-T}(a). This behaviour is summarised in  Fig.~\ref{fig:susz-T}(b) which shows the AF, paramagnetic and FM T=0 longitudinal susceptibilities as function of $\xi$. Asymptotically, for very large $\xi$ they approach zero as in the semiclassical magnets. We note that in the latter an intermediate value for the $T=0$ susceptibility may occur for {\it polycrystalline} case where an {\it averaging} over longitudinal and transverse susceptibility occurs.\\

We now turn to the dynamical properties of the three models as evidenced by the magnetic exciton dispersions, their relation to the induced moments and their critical behaviour. As mentioned before we treat only the FM case in order to avoid the complications with the unit cell doubling in AF case. The SSM case is presented in Fig.~\ref{fig:disps-q}. It shows the single exciton branch for a $\xi_s=1.5$ above the QCP for various temperatures. As the latter approaches the $T_m$ for induced FM moment formation a complete softening $\omega(\bq_m,T)\rightarrow 0$ at the incipient ordering vector $\bq_m =0$ ($\Gamma$ -point) is observed (red curve). Below $T_m$ the soft mode at $\Gamma$  shows rapid re-hardening since there is no continuous rotation symmetry that protects it as in the SDM case. We also include the intensity $\hat{R}(\bq)$ (dash-dotted lines) which show a pronounced peak at the soft mode position and singular behaviour at $T_m$ but modest variation elsewhere in the BZ.\\

The SDM model displays a more intricate behaviour of excitation spectrum presented in Fig.~\ref{fig:dispd-q} as spectral intensity plot for three different temperatures in the induced moment phase. There are now three possible modes appearing. Two of them correspond to dispersive excitations from the singlet ground state to the split doublet states which are different for  polarisation parallel and perpendicular to the induced moment  ($\omega_x, \omega^+_y$). The remaining transverse low energy mode $\omega^-_y$  originates from the thermally activated excitation between the split doublet components. As one can see close to the $\Ga$-point  it hybridises with the dispersive high energy transverse $\omega_y^+$  mode and outside the hybridisation region is almost dispersionless and rapidly looses intensity. Therefore the soft mode at $\Ga$ corresponds to a hybridised singlet-doublet and thermally activated doublet-doublet mode.  It is the  Goldstone mode of the ordered phase since it is zero at the FM ordering vector ($\Gamma$-point) for all temperatures. On approaching $T_m$ from below the $\omega^-_y$ mode is pushed to zero energy due to the vanishing $\Gamma_4$ splitting and essentially becomes a quasi-elastic excitation around the $\Gamma$-point, For $T>T_m$ (not shown) only the fully gapped two high energy $\omega_x,\omega_y$ branches remain which are then degenerate throughout the BZ.

The hybridisation of excitations from the ground state with thermally excited transitions is a fundamental feature of the dynamics of singlet induced moment systems. The evidence for the importance of the latter may also be seen in the non-monotonic temperature dependence of the hybridisation gap $\Delta_h(T)=\omega_y^+(\bq_r)- \omega_y^-(\bq_r)$   between the two modes with $\bq_r(T)$ denoting the crossing wave vector as plotted in Fig.~\ref{fig:dhyb-T}(a):  At low temperature $\Delta_h(T)$  vanishes because of doublet depopulation, then achieves a maximum for
an intermediate $T/T_m\simeq 0.8$ and drops steeply to zero when the soft mode at $T_m$ is approached and the doublet splitting tends to zero. The corresponding position $|\bq_r(T)|$ of the hybridisation gap first remains flat at the low-T value given below Eq.~(\ref{eq:dsoft0})  and then also drops to zero, i.e., approaches the $\Ga$-point.
Furthermore the behaviour of the three $\bq=0$ modes as function of temperature is shown in (b). Above T$_m$ the two modes of x,y polarisation are degenerate and they become soft modes at $T_m$. While $\omega_y^-(0)$ remains soft for all $T<T_m$ the other  transverse and longitudinal modes show a re-hardening below $T_m$.\\

Finally in the STM case we encounter another interesting and important feature of the dynamics in singlet ground state magnetism. In this model we discuss only the longitudinal modes (in the cubic case the moment may be oriented along any axis chosen as z here). Its temperature dependent dispersion is presented in Fig.~\ref{fig:dispt-q}(a). First it shows the usual softening at the ordering vector ($\Gamma$-point) designated by the blue arrow. However, unlike in the previous case it does not come down to zero energy  but is arrested at a finite energy at the transition temperature. Below $T_m$ it immediately re-hardens (red arrow). The reason becomes clear when we consider  Fig.~\ref{fig:dispt-q}(b) where the $\Gamma$- longitudinal mode is plotted in the paramagnetic and induced moment regime. If we extrapolate the paramagnetic mode energy beyond $T_m$ it would indeed become soft at the approximate  transition temperature $T_m^0<T_m$ that does not contain the effect of the Curie type contributions from the excited triplet with matrix element $m'_t=1$. Because of their presence the actual transition occurs at a higher temperature $T_m$ where the mode has not yet become soft. And below $T_m$ it re-hardens immediately due to the  effect of the molecular field on the singlet-triplet splitting. This behaviour is presented in more detail in Fig.~\ref{fig:softmode-t}. The left panel (a) shows that the arrested $\omega(0,T_m)$ increases rapidly linearly with the ratio of elastic (intrinsic $\Gamma_4$) to inelastic ($\Ga_1-\Ga_4$) matrix elements whereas in (b) the square- root increase close to the QCP from numerical evaluation and the approximate expression in Eq.~(\ref{eq:softapprox}) is observed.  \\

This arrested softening effect is not constrained to only the Curie-type contributions in the STM. If, for any of the singlet ground state models discussed here the influence of even higher CEF multiplets are considered there will always be such terms which increase the transition temperature to a value $T_m$ before the softening of the mode connected with the nondiagonal induced  order from the inelastic transition is achieved. Therefore, Figs.~\ref{fig:dispt-q},\ref{fig:softmode-t} illustrate in the simplest possible model what will happen quite generally to the temperature dependence of magnetic excitons: There will be some softening observed but it is hardly ever complete, even on the RPA level due to the influence of thermally excited Curie type contributions from higher level that increase the transition temperature. To achieve a maximum softening effect at $T_m$ one should have i) A control parameter $\xi$ close to the QCP such that at $T_m\ll\De$ all thermally excited Curie contributions enhancing $T_m$ are exponentially suppressed and ii) the diagonal matrix elements of the first excited multiplet should be small. Furthermore beyond RPA one has to consider 
scattering effects of magnetic excitons which in principle will lead to a broadening of the approximate soft mode into a quasielastic peak before the transition occurs~\cite{jensen:91}.\\

{\BLU The aspect of the influence of hyperfine coupling on induced order  has not been included in our discussion.
As mentioned in the Introduction where we have cited related references  to it this becomes particularly important
close to the quantum critical point (below as well as above $\xi_c=1$). In this restricted region it may indeed be highly interesting how the thermodynamic properties are modified, e.g. the nuclear contributions to the specific heat jump close to critical condition and also how the combined nuclear moments and 4f excitation spectrum evolves, in particular for the SDM model with its peculiar hybridised soft mode structure.}

\section{Summary and conclusion}

We have undertaken a comparative investigation of quantum critical and thermodynamic properties as well as magnetic excitations in the most typical singlet ground state quantum magnets. These models are realised approximately in various non-Kramers f-electron compounds  containing  frequently Pr or U  whose f-shells have total angular momentum  J = 4 or other integer values split into CEF multiplets with a nonmagnetic singlet ground state. In this case the conventional quasi- classical magnetic order of Kramers degenerate ground state compounds is not possible. The order can only appear via spontaneous superposition with excited CEF states belonging to singlet, doublet or triplet caused by intersite exchange and facilitated by a nondiagonal matrix element of angular momentum operators between ground and excited states. In this mechanism the moment and its ordering appear simultaneously.\\

This type of quantum magnetism is governed by a control parameter characterising the relative strength of intersite exchange to CEF splitting. Below a critical value for this parameter the ground state is paramagnetic and a spontaneous induced moment appears when the control parameter crosses the quantum critical point. The behaviour of the induced moment and the transition temperature  near the QCP show a logarithmic singularity meaning a rather sudden appearance of the moment and a finite ordering temperature.\\

There are several distinctions in the thermodynamic properties of singlet ground state magnets as compared to the quasi-classical ones with degenerate ground state: Firstly, in the latter the specific heat discontinuity at the ordering temperature is an interaction independent constant while in the former it strongly depends on the control parameter characterising the interaction strength to splitting ratio. Therefore the size of the specific heat jump tends to zero when approaching the QCP from above, concomitant with the vanishing size of the ordered induced saturation moment. {\BLU Secondly for control parameter moderately above the critical one the induced moment  will still be much less than the paramagnetic high temperature effective moment. Furthermore the saturation moment and transition temperature in the vicinity of the QCP vary steeply with the control parameter which may be identified by application of pressure that modifies the distance to the QCP.}
Thirdly in the ordered state of quasi-classical magnets the longitudinal susceptibility tends to zero at low temperatures while in the singlet ground state quantum magnets it has generally a nonzero value that depends on the control parameter and reaches zero only for very large values corresponding to a quasi-degenerate CEF level system. {\BLU
These major differences should be helpful criteria in distinguishing singlet ground state quantum magnetism  from 
quasi-classical magnetism.}\\

{\BLU
The thermodynamic properties are qualitatively similar, though quantitatively different, for the three singlet ground state models investigated, In particular this applies to the size of specific heat jumps relative to the underlying Schottky-type anomaly in the three models because of the considerable difference in entropy release for the various excited multiplet degeneracies. The distinction is even more pronounced for the dynamical properties as evidenced by the dispersion of magnetic exciton modes and its dependence on temperature and control parameter. }As opposed to magnons in the quasi-classical case the magnetic excitons of singlet ground state magnets appear already in the paramagnetic phase as dispersive collective CEF excitations but change their quantitative appearance below the ordering temperature.\\

{\BLU The main mode characteristics and their distinction for the three models may be summarised in the following way:} Firstly in the SSM with only one branch there is a soft mode at the incipient ordering vector at the transition temperature which then re-hardens again due to the Ising-character of the model. Secondly for the SDM several modes appear which become nondegenerate below the ordering temperature, one of the transverse modes turns into a Goldstone mode for all temperatures below $T_m$ while the other two (longitudinal and transverse) modes show again a typical re-hardening from the soft mode below the transition. Finally in the the STM an important effect of the quasi-Curie type contributions to the static susceptibility originating from the triplet states can be identified: Their influence on the ordering temperature leads to an arrested, only partial exciton mode softening at the true transition temperature with a further re-hardening of the mode energy below it. Such partial softening is commonly observed  in real singlet ground state compounds because of the influence of higher lying multiplets not contained in the two-multiplet simplified models analysed here.

\begin{acknowledgments}
The authors thank Burkhard Schmidt for helpful discussions on the CEF symmetry properties.
A.A. acknowledges
the financial support from the German Research Foundation
within the bilateral NSFC-DFG Project No. ER 463/14-1.
\end{acknowledgments}

\appendix

\section{CEF singlet ground state models}
\label{app:CEF}

Here we give possible realisations (among many others) of the three singlet ground state models investigated.
We focus on $J=4$ representations relevant for Pr and U- compounds. The first two models are for uniaxial symmetry 
the last one for cubic symmetry. The notations for CEF states on the left are used in the main text while those to the right are in free ion $|J,M\rangle =|M\rangle$ notation with the point group representation indicated. Within these state spaces the magnetic dipolar operators have the form given in Sec.\ref{sec:model}.
\be
\bl
&\mbox{SSM:} 
\\
&|0\ra = \cos\theta |0\ra  +\sin\theta\frac{1}{\sqrt{2}} (|+4\ra+|-4\ra) \;\;(\Gamma^{(1)}_1)
\\
&|1\ra =\frac{1}{\sqrt{2}}(|+4\ra-|-4\ra) \;\; (\Gamma_2)
\quad\quad\quad
\quad\quad\quad
\quad
\el
\ee
This Ising-type singlet-singlet model is relevant for Pr - compounds with uniaxial symmetry (e.g. $D_{4h}$) 
 and has in particular been confirmed for for URu$_2$Si$_2$~\cite{sundermann:16,marino:23b} from spectroscopic results.
 In this case the CEF mixing angle $\theta$ is close to $\pi/2$.
 The Ising moment operator is $J_z=m_sS_x$ in pseudo spin presentation within this subspace (Eq.~(\ref{eq:Jssm})) with
$ m_s=8\sin\theta$. We finally remark that a singlet-singlet level scheme in a uniaxial symmetry cannot support an xy-type SSM  and
an inspection of the point group multiplication tables leads to the conclusion that this is forbidden for any symmetry.
\be
\bl
&\mbox{SDM:}
\\
&|0\ra = |0\ra \;\;(\Gamma_1)
\\
&|1+\ra =|+1\ra \;\; (\Gamma^+_6)
\\
&|1-\ra =|-1\ra \;\; (\Gamma^-_6)
\quad\quad\quad
\quad\quad\quad
\quad\quad\quad
\quad\quad
\el
\ee
This xy-type singlet- doublet model was proposed for hexagonal (D$_{6h}$) UGa$_2$ from spectroscopic results~\cite{marino:23a}. The moment operators $J_\al=m_dS_\al$ (Eq.~(\ref{eq:Jsdm})) $(\al=x,y)$ are then
defined with the matrix element $m_d=\sqrt{10}$. 
\be
\bl
&\mbox{STM:}
\\
&|\psi_0\ra = \frac{\sqrt{21}}{6}|0\ra +\frac{\sqrt{30}}{12}(|+4\ra+|-4\ra) \;\;(\Gamma_1)
\\
&|\psi_1\ra =-(c|-3\ra+d|+1\ra) \;\; (\Gamma^+_4)
\\
&|\psi_2\ra =\frac{1}{\sqrt{2}}(|+4\ra-|-4\ra)\;\; (\Gamma^0_4)
\\
&|\psi_3\ra =c|+3\ra+d|-1\ra \;\; (\Gamma^-_4)
\el
\ee
where $c=\sqrt{\frac{1}{8}}$ and  $d=\sqrt{\frac{7}{8}}$. This $J=4$ singlet-triplet model is appropriate
for cubic $(O_h)$ Pr- and U- compounds~\cite{lea:62,koga:06}. It leads to the $J_z$ moment operator defined
in Eq.~(\ref{eq:Jstm}) with $m_t=\frac{4}{3}\sqrt{15}$ and $m'_t=1$ fixed by symmetry for $J=4$.  For larger integer $J$ the $\Gamma_{1,4}$ occur multiple times and then $m_t, m'_t$ depend on CEF potential parameters.

\section{Generalized Schottky anomaly}
\label{app:schottky}

In this Appendix we give the specific heat for a singlet ground state - N-fold degenerate excited state multiplet two level system in the paramagnetic state. In reality a maximum of $N=3$ is possible for CEF states in cubic point group symmetry, except when further accidental degeneracy occurs. For the general case we obtain
\be
C_V(T)=\frac
{4N\bigl(\frac{\De}{2T} \bigr)^2 }
{\bigl[(N+1)\cosh\bigl(\frac{\De}{2T} \bigr)-(N-1)\sinh\bigl(\frac{\De}{2T} \bigr)\bigr]^2}
.
\label{eq:NCV}
\ee
For the SSM case (N=1) one obtains the common Schottky anomaly given in the second line of Eq.~(\ref{eq:sCV}) and 
for the SDM (N=2) and STM (N=3) cases the explicit expressions are given in Eqs.~(\ref{eq:dCV},\ref{eq:tCV}), respectively. The low and high temperature limits of the above general expression are obtained as
\be
\bl
 C_V(T)\simeq
\left\{
\begin{array}{l l}
N\bigl(\frac{\De}{T} \bigr)^2e^{-\frac{\De}{T}};\;(T\ll\De)\\[0.3cm]
\bigl(\frac{N}{N+1}\bigr)^2\bigl(\frac{\De}{T} \bigr)^2;\;\;\;\;(T\gg\De)
\end{array}
\right.
.
\el
\ee

\section{Spectral function for the SDM}
\label{app:dspectral}

Here we give the magnetic exciton intensities that define the spectrum of the dynamical magnetic susceptibility of the SDM. Because of the appearance of three exciton branches it is more involved than for the single branch in the SSM case (Eqs.~(\ref{eq:sintens1},\ref{eq:sintens2})). For the trace over cartesian components $R(\bq,\omega,T)=(1/\pi)\sum_\alpha {\rm Im}\chi_{\al\al}(\bq,\omega)$ we obtain the three mode contributions
\be
\bl
R(\bq,\omega,T)
=&
\hat{R}_{xx}(\bq,T)\delta(\omega-\omega_x)+
\hat{R}^+_{yy}(\bq,T)\delta(\omega-\omega^+_y)
\\&+
\hat{R}^-_{yy}(\bq,T)\delta(\omega-\omega^-_y)
\label{eq:dspecdens}
\el\ee
with the mode dispersions $\omega_x(\bq,T)$ and $\omega^\pm_y(\bq,T)$ given in Eqs.~(\ref{eq:xxmode0},
\ref{eq:yymode0}). The intensities of each branch are obtained as
\be
\bl
\hat{R}_{xx}(\bq,T)
=&
\bigl(\frac{\De}{I_e}\bigr)\bigl(\frac{\De}{2\omega_{x\bq}}\bigr),
\\
\hat{R}^+_{yy}(\bq,T)
=&
\\
&\hspace{-1cm}
\bigl(\frac{\De}{I_e}
\bigr)\Bigl[W_{21}^+\frac{\omega^{+2}_{y\bq}-(\De_T^{32})^2}{\omega^{+2}_{y\bq}-(\omega^-_{y\bq})^2}+
W_{32}^+\frac{\omega^{+2}_{y\bq}-(\De_T^{21})^2}{\omega^{+2}_{y\bq}-(\omega^-_{y\bq})^2}\Bigr],
\\
\hat{R}^-_{yy}(\bq,T)
=&
\\
&\hspace{-1cm}
\bigl(\frac{\De}{I_e}
\bigr)\Bigl[W_{21}^-\frac{(\De_T^{32})^2-(\omega^-_{y\bq})^2}{\omega^{+2}_{y\bq}-(\omega^-_{y\bq})^2)^2}+
W_{32}^-\frac{(\De_T^{21})^2-\omega^{-2}_\bq}{\omega^{+2}_{y\bq}-(\omega^-_{y\bq})^2}\Bigr],
\label{eq:dbranchspec}
\el
\ee
where $\De^{21}=\fs(\De_T+\De)$, $\De^{32}=\fs(\De_T-\De)$ are excitation energies between the molecular field
states and $f^{12}_d=p_1-p_2$, $f^{23}_d=p_2-p_3$ the corresponding occupation differences.  Furthermore  we introduced weight coefficients defined by
\be
\bl
W_{21}^\pm&=&\frac{1}{4}(1+\frac{1}{\hDe_T})(\hDe_T+1)\frac{\De_0f_d^{12}}{2\omega_{y\bq}^\pm},
\\
W_{32}^\pm&=&\frac{1}{4}(1-\frac{1}{\hDe_T})(\hDe_T-1)\frac{\De_0f_d^{23}}{2\omega_{y\bq}^\pm}.
\label{eq:dweight}
\el
\ee
An example of the spectral function $R(\bq,\omega,T)$ for SDM is given in Fig.~\ref{fig:dispd-q} (using a finite broadening) and discussed in Sec.\ref{sec:discussion}.
\bibliography{References}

\end{document}